\documentclass[camera,letterpaper,nomarginnotes,nonarrowgutter,hyphens]{jpaper}

\usepackage[sort,compress]{cite}
\usepackage{amsmath,amssymb,amsfonts}
\usepackage{algorithmic}
\usepackage{graphicx}
\usepackage{textcomp}
\usepackage{xcolor}
\usepackage{fancyhdr}
\usepackage{hyperref}

\usepackage[font=footnotesize]{subcaption}

\usepackage{setspace}

\usepackage[compact]{titlesec}

\usepackage[nolessnomore, italic]{mathastext}
\usepackage[T1]{fontenc}
\usepackage{multirow}
\usepackage{hhline}
\usepackage[normalem]{ulem}
\usepackage{indentfirst}
\usepackage{footmisc}

\definecolor{darkpink}{rgb}{0.88, 0.28, 0.54}
\definecolor{forestgreen}{rgb}{0.0, 0.27, 0.13}
\definecolor{amber}{rgb}{1.0, 0.49, 0.0}

\titlespacing\section{0pt}{4pt plus 2pt minus 2pt}{4pt plus 2pt minus 2pt}
\titlespacing\subsection{0pt}{4pt plus 2pt minus 2pt}{4pt plus 2pt minus 2pt}
\titlespacing\subsubsection{0pt}{4pt plus 2pt minus 2pt}{4pt plus 2pt minus 2pt}
\makeatletter
\g@addto@macro{\normalsize}{%
  \setlength{\abovedisplayskip}{5pt plus 0.5pt minus 1pt}
  \setlength{\belowdisplayskip}{5pt plus 0.5pt minus 1pt}
  \setlength{\abovedisplayshortskip}{3pt}
  \setlength{\belowdisplayshortskip}{3pt}
  \setlength{\intextsep}{3pt plus 1pt minus 1pt}
  \setlength{\textfloatsep}{3pt plus 1pt minus 1pt}
  \setlength{\skip\footins}{3pt plus 1pt minus 1pt}}
\makeatother

\definecolor{dollarbill}{rgb}{0.52, 0.73, 0.4}

\hypersetup{
citecolor=blue,
linkcolor=blue,
urlcolor=blue
}

\fancypagestyle{firstpage}
{

    \fancyfoot[C]{\thepage}
}

\newcommand*\circled[1]{\kern-2.5em%
  \put(0,3){\color{white}\circle*{10}}\put(0,3){\circle{8}}%
  \put(-2,0.5){\color{black}#1}~~}
\usepackage{enumitem}

\newcommand{\circlegf}[1]{\tikz[baseline=(char.base)]{\node[shape=circle,draw,inner sep=0pt,fill=white, text=black] (char) {#1};}}

\newcounter{takeawaycounter}

\newcommand{\takeaway}[1]{
    \stepcounter{takeawaycounter}
    \begin{samepage}
    \colorbox{gray!40}{\textbf{Key Takeaway \thetakeawaycounter:}} \textit{#1}
    \end{samepage}
}

\newcounter{observationcounter}

\newcommand{\observation}[1]{
    \stepcounter{observationcounter}
    \begin{samepage}
    \colorbox{gray!40}{\textbf{Key Observation \theobservationcounter:}} \textit{#1}
    \end{samepage}
}

\newcounter{recomcounter}

\newcommand{\recommendation}[1]{
    \stepcounter{recomcounter}
    \begin{samepage}
    \colorbox{gray!40}{\textbf{Recommendation \therecomcounter:}} \textit{#1}
    \end{samepage}
}

\def\BibTeX{{\rm B\kern-.05em{\sc i\kern-.025em b}\kern-.08em
    T\kern-.1667em\lower.7ex\hbox{E}\kern-.125emX}}

\title{PIMDAL: Mitigating the Memory Bottleneck in Data Analytics\\
using a Real Processing-in-Memory System}

\author{
Manos Frouzakis\textsuperscript{a}~\qquad
Juan Gómez-Luna\textsuperscript{b}~\qquad
Geraldo F. Oliveira\textsuperscript{a}~\qquad \\
Mohammad Sadrosadati\textsuperscript{a}~\qquad
Onur Mutlu\textsuperscript{a}\vspace{10pt}
\\
\textsuperscript{a}~\emph{ETH Zürich} \qquad
\textsuperscript{b}~\emph{NVIDIA Research}
\vspace{5pt}
}

\begin{document}

\maketitle
\thispagestyle{firstpage}

\begin{abstract}

Database Management Systems (DBMSs) are crucial for efficient data management and analytics, and are used in several different application domains. Due to the increasing volume of data a DBMS deals with, current processor-centric architectures (e.g., CPUs, GPUs) suffer from \emph{data movement bottlenecks} when executing key DBMS operations (e.g., selection, aggregation, ordering, and join). This happens mostly due to the limited memory bandwidth between compute and memory resources.

Data-centric architectures like Processing-in-Memory (PIM) are a promising alternative for applications bottlenecked by data, placing compute resources close to where data resides. Previous works have evaluated using PIM for data analytics. However, they either do not use real-world architectures or they consider only a subset of the operators used in analytical queries. This work aims to fully evaluate a data-centric approach to data analytics, by using the real-world UPMEM PIM system. To this end we first present the \textit{PIM Data Analytics Library (PIMDAL)}, which implements four major DB operators: \texttt{selection}, \texttt{aggregation}, \texttt{ordering} and \texttt{join}. Second, we use hardware performance metrics to understand which properties of a PIM system are important for a high-performance implementation. Third, we compare PIMDAL to reference implementations on high-end CPU and GPU systems. Fourth, we use PIMDAL to implement five TPC-H queries to gain insights into analytical queries.

We analyze and show how to overcome the three main limitations of the UPMEM system when implementing DB operators: (I)~low arithmetic performance, (II)~explicit memory management and (III)~limited communication between compute units. Our evaluation shows PIMDAL achieves $3.9 \times$ the performance of a high-end CPU, on average across the five TPC-H queries.

\end{abstract}

\begingroup\small\noindent\raggedright\textbf{Source Code Availability:}\\
The source code has been made available at \url{https://github.com/CMU-SAFARI/PIMDAL}.
\endgroup

\section{Introduction}

Database Management Systems (DBMSs) \cite{Databases95} provide a standardized interface to manage large amounts of data \cite{DBMS}. They play a key role for data analytics in several application domains, including commerce \cite{AnalyticDB19}, machine learning \cite{MLSpark16}, and medicine \cite{MedDB96}. 
However, due to the volume of data DBMSs process, current processor-centric architectures (e.g., CPU, GPUs, and FPGAs) suffer from \emph{data movement bottlenecks} when executing key DBMS operations \cite{Mondrian17, Polynesia21}. Thus, the performance of database operators are often \emph{bound} by accesses to off-chip main memory (see Section~\ref{goal}). Processor-centric architectures try to circumvent the data movement bottleneck problem by employing a variety of solutions, for example, (I)~physically placing hardware components closer together~\cite{FPGAInterconnect21}, or (II)~leveraging high-bandwidth memory devices~\cite{A100, Ultrascale} (e.g., High Bandwidth Memory (HBM)~\cite{HBMSpec13}). Although such solutions can accelerate database analytics \cite{GPUDBStudy20, FPGADataAnalytics20}, they do \emph{not} \emph{fundamentally} solve the issue caused by off-chip memory accesses. 

In contrast to processor-centric architectures, data-centric architectures, such as \textit{processing-in-memory} (PIM) systems~\cite{PIMPrimer23, PIMcase21, MLPNM21, stone1970logic, elliott1992computational, kogge1994execube, gokhale1995processing, patterson1997case, ghose2019arxiv, Computedram20, levy.microelec14, kvatinsky.tcasii14, kvatinsky.iccd11, gaillardon2016plim, Draper:2002:ADP:514191.514197, eckert2018neural, fujiki2019duality, kang.icassp14, Ambit17, seshadri2016buddy, Rowclone18, angizi2019graphide, kim.hpca18, xin2020elp2im, li.micro17, ali2019memory, angizi2019dna, shafiee2016isaac, bhattacharjee2017revamp, kim.bmc18, ahn.pei.isca15, morad.taco15}, can mitigate the main memory bottleneck in data analytics~\cite{Prim22} by performing computation where the data resides~\cite{PIMPrimer23}, i.e., inside the main memory. 
Recently, industry has announced several PIM architectures, including general-purpose PIM architectures such as the UPEM PIM system~\cite{UPMEM} and application-specific PIM architectures such as Samsung Aquabolt-XL HBM2-PIM~\cite{Aquabolt22} and SK Hynix AiM~\cite{AiM21}.
In this work we focus on the UPMEM PIM architecture, since it is the only commercially available PIM architecture currently on the market. The UPMEM architecture consists of general-purpose in-order cores, placed close to DRAM banks.

Previous works have shown the potential of the UPMEM PIM system for accelerating a subset of DB operators \cite{Prim22, JoinDIMM23, Poseidon23}. None of these works however, have evaluated full analytical queries, for example ones commonly encountered in practice. The main \textbf{goal} of this work is to implement full analytical queries on the UPMEM system to fully characterize data-centric computing for data analytics.

The \textbf{key challenge} in achieving this goal is that current database designs cannot be straightforwardly ported to PIM, since they are designed for conventional architectures. The UPMEM system introduces new limitations, namely low arithmetic performance, explicit memory management and limited communication across compute units. In order to address these limitation we explore different algorithmic designs based on the commonly used DB operator building blocks \emph{sorting} and \emph{hashing}. We also demonstrate the usage of advanced UPMEM SDK communication functionality to solve the challenge posed by communication.

We implement the \textbf{four DB operators} \texttt{selection}, \texttt{aggregation}, \texttt{ordering} and \texttt{join}, which are building blocks for the TPC-H benchmark \cite{TPC} queries,
a widely used benchmark for the evaluation of the performance of commercial DBMSs. We use five queries from the TPC-H benchmark to compare PIMDAL to CPU and GPU implementations using \textit{PyArrow} \cite{PyArrow} and \textit{cuDF} \cite{cudf}, respectively. PIMDAL outperforms the CPU implementation by $3.9 \times$ on average for the five TPC-H queries. Compared to a high-end GPU it achieves a speedup of $2.2 \times$ and $3.3 \times$ in two of the queries. The other queries require more complex communication between compute units, which we find to be a key weakness of current PIM systems.

\begin{itemize}[leftmargin=3mm,itemsep=0mm,parsep=0mm,topsep=0mm]
    \item We implement the database operators \texttt{selection}, \texttt{aggregation}, \texttt{sorting}, and \texttt{join}. They are relevant in commercial DBMS as shown by their use in the TPC-H benchmark. 
    \item We evaluate the operator design using selected queries from the TPC-H benchmark targeting database management systems used in practice.
    \item We provide recommendations for implementation and also architecture improvements to better support data analytics.
\end{itemize}
 
\section{Background and Motivation}

\subsection{Database Management Systems} 
\label{DBMS}

DBMSs manage and provide efficient access to large amounts of data. The development of DBMS started in the late 1950s~\cite{DBMS}, even before the invention of the microprocessor. The goal was to introduce generalized routines for efficiently operating on files and data. Databases have evolved together with computer systems, for example with the introduction of main memory databases \cite{MMDB}. Larger main memory sizes enable storage of the entire database within the main memory, delivering savings in terms of execution time and energy consumption. However, main memory database operators still suffer from data movement bottlenecks \cite{Prim22}.

In this work, we consider relational databases using structured data \cite{RelationalDatabase}, which consist of tables where each data point is a row. A row is a tuple, with each entry belonging to a set, such that a column of the table consists of elements of a set. Database (DB) operators define operations that are performed on the rows or columns of one or multiple tables. Table~\ref{tab:operators} shows the DB operators we consider in this work.

\begin{table}[h]
    \centering
    \caption{Key operators supported by many DBMS.}
    \resizebox{\linewidth}{!}{
    \begin{tabular}{|l||p{6cm}|}
        \hline
         DB Operator & Function \\
        \hline
        \hline
        \textbf{\texttt{Selection}} & Filters tuples in a database based on a predicate on some columns. \\
        \hline
        \textbf{\texttt{Aggregation}} & With \texttt{aggregation} we refer to \texttt{grouping} with \texttt{aggregation} \cite{Aggregate82}. Rows are grouped together, based on the values of specified columns. The rest of the columns are either removed or aggregated with an aggregation function. \\
        \hline
        \textbf{\texttt{Ordering}} & Orders the tuples based on some columns.\\
        \hline
        \textbf{\texttt{Join}} & Concatenates the rows of one table (\textit{inner relation}) to the rows of another table (\textit{outer relation}) based on key values \cite{RelationalDatabase}. \\
        \hline
    \end{tabular}
    }
    \label{tab:operators}
\end{table}

\subsection{Processing-In-Memory Architectures} 
\label{arch}

Processing-in-memory (PIM) architectures alleviate the data movement bottleneck caused by off-chip memory accesses \cite{PIMPrimer23, PIMcase21, MLPNM21, stone1970logic, elliott1992computational, kogge1994execube, gokhale1995processing, patterson1997case, ghose2019arxiv, Computedram20, levy.microelec14, kvatinsky.tcasii14, kvatinsky.iccd11, gaillardon2016plim, Draper:2002:ADP:514191.514197, eckert2018neural, fujiki2019duality, kang.icassp14, Ambit17, seshadri2016buddy, Rowclone18, angizi2019graphide, kim.hpca18, xin2020elp2im, li.micro17, ali2019memory, angizi2019dna, shafiee2016isaac, bhattacharjee2017revamp, kim.bmc18, ahn.pei.isca15, morad.taco15}. Processing-near-memory (PNM) \cite{ipim20, FAFNIR21, WeatherPNM21, GradPIM21, Polynesia21, NATSA20, NAPEL19, Chameleon16, JAFAR15, NDA15, gao.pact15, HRL16, TOM16, Neurocube16, kim.sc17, nai2017, pugsley2014ndc, zhang.hpdc14, zhu2013, akin2015, Tetris17} is one type of PIM, that places processing elements (PEs) close to the main memory. As a result, PNM systems can access the memory with a much higher parallelism than through a shared memory bus, used in conventional processors (CPU, GPU).
UPMEM PIM \cite{UPMEM} for example, accelerates general-purpose, memory-bound workloads with simple, RISC-style cores near DRAM memory banks. 
Samsung Aquabolt HBM2-PIM \cite{Aquabolt22} and SK Hynix GDDR6-AiM \cite{AiM21} both aim at accelerating machine learning and artificial intelligence workloads, using SIMD PEs that execute multiply-and-accumulate operations.

The UPMEM PIM architecture is implemented on standard DDR4-2400 DRAM technology \cite{UPMEM}. It is designed to interact with a host CPU by replacing DRAM DIMMs. It consists of a set of PIM-cores with private DRAM, called DPUs. An UPMEM PIM rank consists of 8 DRAM banks, which in turn consist of 8 PIM-cores each. They are controlled at the granularity of a rank, by sending commands and data over the DRAM bus. Therefore, when accessing a rank the host system communicates with up to 64 PIM-cores at once.

Each UPMEM PIM-core has exclusive access to a \textbf{64 MiB DRAM} (MRAM), \textbf{64 KiB scratchpad memory} (WRAM) and \textbf{24 KiB of instruction memory} (IRAM) \cite{Prim22}. The MRAM stores the data a PIM-core processes. However, it cannot directly do this in MRAM, it needs to first load it into scratchpad memory (SPM), the WRAM, similarly to caches in conventional CPUs. In contrast to CPUs, these loads have to be performed by the PIM software, as we explain later. Each PIM-core can independently execute its own program.

Ideally, future PIM architectures would replace the main memory in computer architectures. However, current CPU memory controllers are not equipped to deal with PIM memory, due to its different data layout than commodity DRAM \cite{Prim22}. Instead, all accesses from the host have to be managed in the host software, copying data to pointers defined in the PIM code \cite{UPMEMTransfers}. There are four types of data transfers:
(I)~\textbf{Serial transfers} copy a contiguous memory section from or to one specific PIM-core.
(II)~\textbf{Broadcast transfers} copy the same contiguous memory to a set of PIM-cores.
(III)~\textbf{Parallel transfers} copy from/to a set of aligned, contiguous memory regions each to/from a different PIM-core in parallel.
(IV)~\textbf{Scatter/gather transfers} copy disjoint memory sections from or to different PIM-cores in parallel. PIM-cores cannot directly communicate together, only through data transfers over the host. This can lead to costly host data accesses, that can potentially become a system bottleneck~\cite{Prim22, JoinDIMM23, DIMMLink23}. In the current UPMEM architecture, the DRAM of a PIM-core is inaccessible to the host for data transfers during PIM execution. However, it is possible to schedule transfers and execution asynchronously for different ranks, so that the data on idle ranks can be accessed while others are executing. This can optimize bandwidth utilization by spreading out transfers.

Compared to the out-of-order cores in an x86 CPU, PIM cores use a \textbf{in-order pipeline} \cite{Prim22} with a \textbf{custom, RISC-like ISA}. An important metric for the performance of a processor is the instructions per cycle (IPC). CPUs can execute multiple instructions simultaneously, having a maximum IPC higher than 1. The in-order pipeline in contrast can only reach a maximum IPC of 1. In practice, the achieved IPC depends on how fast the cores can be fed with data. This is why PIM trades off arithmetic performance for memory bandwidth. We will be using the IPC to evaluate how efficiently compute resources are used. To reach peak IPC, corresponding to full pipeline occupancy, PIM-cores rely on multi-threading. All threads share the same memory space, operating using the multiple instructions, multiple data (MIMD) paradigm. Each thread can execute its own code section, which can be assigned using the thread ID. Due to very limited parallelism in the pipeline, instructions from the same thread have to be dispatched at least 11 cycles apart \cite{Prim22}. \textbf{This means full pipeline occupancy requires at least 11 working threads}. Above 11 threads the performance improvement depends on the workload. A thread can mitigate stalls in other threads, but also interfere with them. In terms of arithmetic operations, \textbf{only 32 bit integer addition and subtraction are natively supported} by the hardware \cite{Prim22, UPMEMML23, Simplepim23}. Multiplication, division and other datatypes are emulated in software (e.g. 64 bit addition is compiled to two 32 bit additions). To access MRAM, a direct memory access (DMA) instruction has to be used, that copies a variable-length segment from or to the SPM \cite{Prim22}. The accessed addresses have to be aligned to 8 bytes and data transfer sizes a multiple of 8 bytes.
MRAM latency has a constant plus a variable component depending on access size \cite{Prim22} ($\alpha \cdot size + \beta$), meaning \textbf{accessing larger arrays better amortizes the cost of memory accesses}.

Table \ref{tab:architectures} shows key characteristics of the PIM architecture compared to an \textit{Intel Xeon 6226R Gold CPU}. While the arithmetic performance of the CPU is higher, the PIM system has magnitudes higher bandwidth.

\begin{table}[h]
    \centering
    \caption{Key characteristics of different architectures.}
    \resizebox{\linewidth}{!}{%
    \begin{tabular}{|c|c|c|c|}
        \hline
         \textbf{Characteristics} & \textbf{DPU} & \textbf{PIM System} & \textbf{CPU} \\
        \hline
        \hline
        \textbf{DRAM Read BW} & 628 MB/s & 1286 GB/s & 79 GB/s \\
        \hline
        \textbf{DRAM Write BW} & 633 MB/s & 1296 GB/s & 79 GB/s \\
        \hline
        \textbf{Cache BW} & 2818 MB/s & 5771 GB/s & 1630 GB/s\\
        \hline
        \textbf{Peak Int Throughput} & 350 MOps & 760 GOps & 1606 GOps \\
        \hline
    \end{tabular}}
    \label{tab:architectures}
\end{table}

\subsection{Motivation and Goal} \label{goal}

\noindent \textbf{Memory-Bound DB Operators.} Main memory accesses in current processor-centric architectures make up a big fraction of the execution time and energy consumption of database operators. To understand how main memory access impact the performance of key DB operations, we calculate the Roofline model~\cite{Roofline09} of the four DB operators \texttt{selection}, \texttt{aggregation}, \texttt{ordering}, and \texttt{join}. The Roofline model indicates whether an application is bound by compute or memory resources. The horizontal lines show if operations are limited by the arithmetic performance of the CPU. The bottom line is the maximal performance using scalar arithmetic instructions. The top line is for using higher performance, vectorized instructions in modern processors, which can be used for some DB operators \cite{zhou2002dbsimd}. The inclined lines on the left show the operations being limited by the data transfers. The bottom line indicates the operations being limited by DRAM bandwidth, when the memory access pattern is inefficient. The top line for the L1 cache shows how effectively an implementation leverages the cache. The x-axis shows the arithmetic intensity in op/byte, which determines what component limits the performance. The y-axis shows the effective performance achieved. We profile DB operator implementations and create the roofline plot using Intel Advisor~\cite{Advisor}.

We observe from the figure that all four DB operators fall into the memory-bound region of the Roofline model (the left side of the intersection between the DRAM bandwidth line and the peak scalar arithmetic performance), which indicates that main memory bandwidth is the primary performance limiter for such applications. The memory access pattern of the \texttt{selection} operator makes most requests go to main memory. It cannot efficiently use the cache and is thus strictly bound by DRAM bandwidth. \texttt{Aggregation} performs some operations on the cache, meaning it is less affected by this, outperforming the DRAM roofline. \texttt{Ordering} moves data around many times, without being able to fully take advantage of the cache. This leads to low arithmetic intensity and low performance. Hash \texttt{join} also moves data around many times, mainly because it partitions the data. This leads to lower arithmetic intensity, but improves performance by better taking advantage of caches. Overall, we conclude that such DB operations can benefit from data-centric architectures, such as the UPMEM PIM system, due to their ability to mitigate memory bottlenecks.

\begin{figure}[ht]
    \centering
    \includegraphics[width=0.8\linewidth]{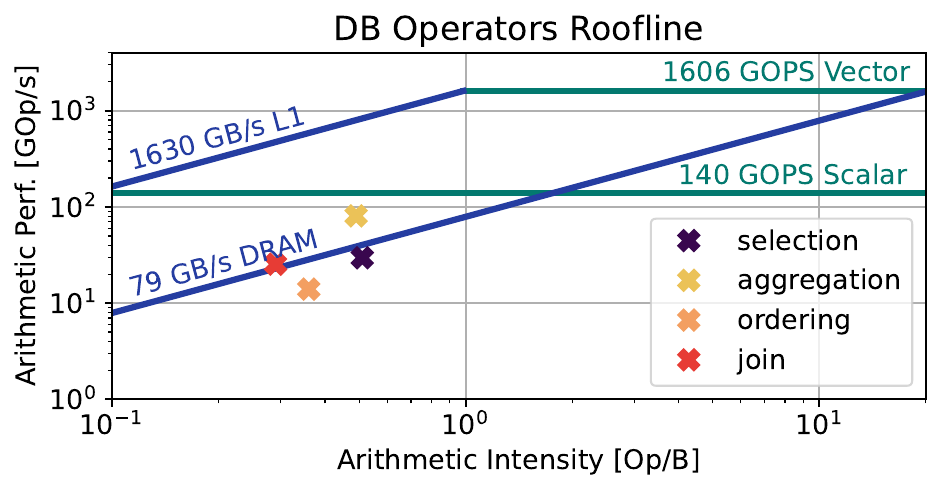}
    \caption{Roofline model of DB operators running on an Intel Xeon Gold 6226R.}
    \label{fig:roofline}
\end{figure}

\noindent \textbf{UPMEM PIM: Challenges \& Limitations.} Even though the UPMEM PIM system can alleviate the memory bottlenecks faced by DB operations, naively implementing such operators in the UPMEM system can lead to subpar performance for three main reasons:
(I)~Data movement (i.e., DRAM to SPM data transfers) has to be explicitly managed by the programmer. Algorithms with non-trivial access patterns, like sorting and hashing can pose a challenge.
(II)~Hash-based DB operators can require multiplication/division, which is costly on the UPMEM system.
(III)~\texttt{Ordering} and \texttt{join} can require communication over the host system, which potentially creates a new bottleneck.

\noindent Additionally, the UPMEM system requires parallelization for optimal performance. The PIM threads need efficient synchronization and resource sharing, which can differ from conventional systems.

Our \textbf{goal} in this work is to provide and evaluate a high performance implementation of TPC-H database queries on the UPMEM PIM system. We aim to show how data analytics can be improved by using a data-centric approach. The key is to implement DB operators on a PIM system, while addressing all the limitations of said system. To this end, we provide PIMDAL, a library that accelerates data analytics using the UPMEM PIM system.

\section{PIMDAL: PIM Data Analytics Library}
\label{implementation}

This section describes the structure of our \textit{PIM Data Analytics Library (PIMDAL)} that accelerates data analytics using UPMEM PIM. PIMDAL addresses the design limitations of the UPMEM \cite{Prim22} system mentioned in Section \ref{goal} to achieve high performance. From previous characterizations of the system for other workloads \cite{Prim22, UPMEMML23, JoinDIMM23, Simplepim23}, the properties in Section \ref{arch} and challenges in Section \ref{goal} we derive two main design principles. The \textbf{first design principle} is to use \textbf{as big, contiguous DMA transfers as possible}. The \textbf{second design principle} is to is to \textbf{use at least 11 PIM-threads}. We evaluate how different workloads behave with threads counts, also higher ones than 11. The two design principle can actually compete against each other, since the total SPM for the data transfers of all threads is limited. We give an in-depth analysis of these principles and the trade-off in the evaluation. PIMDAL consists of an UPMEM PIM and a host system component.

\noindent \textbf{PIMDAL Host Implementation.} PIMDAL is aiming to provide all the necessary functionality to manage data efficiently. The host system plays a crucial role by providing the PIM system with data and communication capabilities. A way to optimize this is to build on the Apache Arrow \cite{ApacheArrow, UPMEMOLAP} framework, that implements a columnar memory format for data analytics, enabling fast data sharing between devices.
This format is particularly useful for data analytic tasks using PIM, as we show in Section~\ref{mem_management}. Our data is initially stored on the host as an Arrow table using the column-store format \cite{Column09}. Data can also be loaded from disk using the Apache Parquet format~\cite{ApacheParquet}.
For query execution the columns are copied to the PIM-cores, where the query is executed and finally the results are copied back to the host to create an Arrow table again. The host also provides efficient data redistribution between PIM-cores for operators requiring it, described in Section \ref{transfers}.

 The main part of our \textbf{UPMEM PIM Implementation} are the four DB operators \texttt{selection}, \texttt{aggregation}, \texttt{ordering} and \texttt{join}. They are implemented using the UPMEM SDK and running on the PIM-cores \cite{UPMEMSDK}. The operators perform their work on tables stored in column-store format \cite{Column09} in the PIM DRAM banks. We explain how we address the fundamental PIM system limitations when implementing these operators in Section \ref{operators}.

\section{PIM Database Operator Implementation} \label{operators}
\subsection{Building Blocks of DB Operators}

We first describe the implementation of two key algorithms in the development of our DB operations on the UPMEM architecture: sorting and hashing. \textbf{Sorting} can be used as a basis to implement ordering, aggregations, and joins in DBMS~\cite{SortHash94, kim2009sort, muller2015cache}. For \texttt{aggregation} we can group the elements we want to aggregate together by sorting them. Sort-merge \texttt{join} uses sorting to bring the two relations into the same order, and then iterates through them simultaneously to join equal keys.
\textbf{Hashing} is another important \textbf{algorithmic building block in a database management system}~\cite{SortHash94, kim2009sort, muller2015cache}. Like sorting, hashing algorithms can be used to implement both \texttt{aggregations} and \texttt{joins}. Hash \texttt{join} works by first storing key-value pairs from the inner relation in a hash table and probing it with the outer relation. For \texttt{aggregations}, hashing is used to map elements with the same keys to the same location, which enables us to aggregate the other columns.

\subsubsection{Sorting Algorithms on UPMEM} \label{sorting}

The UPMEM system is highly parallel as described in Section \ref{arch}. First, we have multiple PIM-cores working independently. Second, each core needs to run threads in parallel, using shared memory and resources. We show two PIM implementations of commonly used sorting algorithms: \textit{Quicksort} and \textit{Mergesort}. We first explain our single PIM-core, parallel implementation of \textit{Quicksort}~\cite{quicksort} and then show how it can be extended to multi-PIM-core sorting. Then we also demonstrate our parallel \textit{Mergesort} implementation.

\begin{figure}[h]
    \centering
    \includegraphics[width=0.8\linewidth]{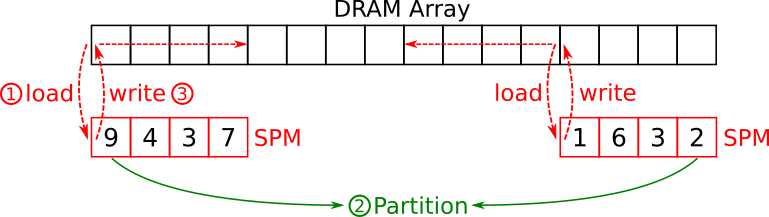}
    \caption{Quicksort PIM implementation.}
    \label{fig:quicksort}
\end{figure}

\textbf{Quicksort} works by repeatedly partitioning an array based on a pivot element \cite{quicksort}. It first chooses a pivot and implicitly places it in the middle of the array, subsequently placing larger elements to the right and smaller ones to the left in the array. Each step produces two partitions of ideally equal size. This operation can be repeated to sort the full array. Our PIM implementation of quicksort executes three main steps (Figure~\ref{fig:quicksort}).
Initially, the input array is contiguously stored in the DRAM bank of a PIM-core. It uses two buffers of size $M$ in the SPM as caches to iterate through the array (\circlegf{1} in Figure~\ref{fig:quicksort}). 
The first buffer ($B_{left}$) loads the first $M$ data elements, and the second buffer ($B_{right}$) loads the last $M$ data elements. 
\circlegf{2} The quicksort algorithm iterates simultaneously through $B_{left}$ to find an element larger and $B_{right}$ to find one smaller than the pivot. These elements are then swapped. In the example we have a pivot value of 5 and swap elements 9 (stored in $B_{left}$) and 2 ($B_{right}$). 
\circlegf{3} If the iteration reaches the end of either $B_{left}$ or $B_{right}$, it stores the content of this buffer back into the DRAM bank of the PIM-core, at the same memory address it was loaded from. Then, it loads the buffer's next $M$ data element portion of the input array, to continue the quicksort execution in step 2.

\begin{figure}[h]
    \centering
    \includegraphics[width=0.9\linewidth]{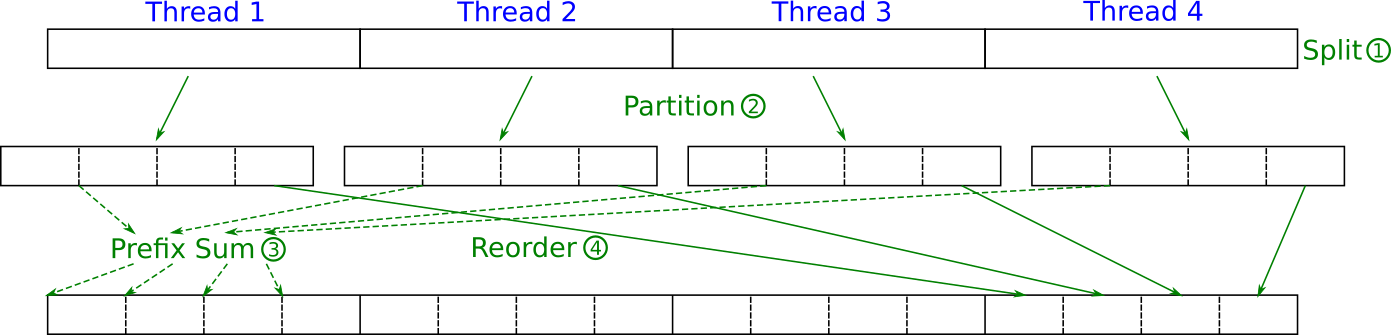}
    \caption{Parallelization technique for quicksort based on sort partitioning.}
    \label{fig:par_quicksort}
\end{figure}

The quicksort algorithm can be naively parallelized by partitioning the initial array using one thread, then using two for the resulting partitions, four for the next ones and so on. This violates the design principle of using at least 11 working threads.
We employ a technique that can better parallelize quicksort to a high number of cores \cite{GPUQuicksort} in four main steps (Figure~\ref{fig:par_quicksort}):
\circlegf{1} Split the initial array into sub-arrays, one for each thread.
\circlegf{2} Let each thread partition its sub-array in place using our quicksort algorithm.
\circlegf{3} Calculate a prefix sum to find the output location for each partition.
\circlegf{4} Write the partitions to a new DRAM array at the calculated locations.

The implementation of quicksort can be either recursive or iterative. Recursion works the same way on the UPMEM system as on other architectures, the only restriction being the stack size. Since the stack occupies space in the SPM \cite{UPMEMStack}, it limits the available buffer size. When implementing quicksort recursively, we found that we have to increase the default stack size to avoid overflows. It can use to up to $38 \si{\kilo\byte}$ out of $64 \si{\kilo\byte}$ available. The iterative implementation is better suited, since it saves SPM space by not having to store the entire function call context.

The concept used for multiple thread partitioning can be applied to \textbf{multi PIM-core sorting} to sort large arrays in parallel. We first partition our array as in single core sorting, but this time we create one partition for each core. These partitions are transferred to the host together with the indices, where they are reordered and copied back to the PIM-cores so that each core now receives a contiguous range of elements. Each PIM-core can sort its data independently, after which the whole data will be sorted.

\begin{figure}[h]
    \centering
    \includegraphics[width=0.8\linewidth]{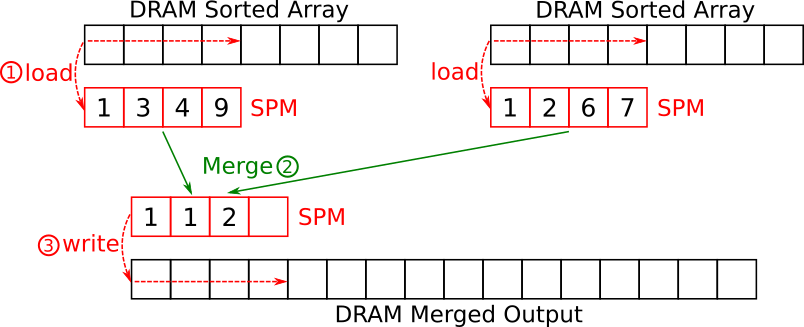}
    \caption{Mergesort PIM implementation.}
    \label{fig:mergesort}
\end{figure}

\textbf{Mergesort} works by repeatedly merging two sorted arrays, always choosing the smaller element at the start of the two arrays. It first splits the initial array into single element chunks that can then be merged. Mergesort has regular, sequential memory access patterns, which make it well suited for implementation on the UPMEM architecture~\cite{Prim22}. Figure~\ref{fig:mergesort} shows one iteration of our implementation using three buffers allocated in the SPM of a PIM-core: two buffers containing the initial sorted arrays, and one buffer for the merged array. The algorithm works in three main steps (Figure~\ref{fig:mergesort}):
\circlegf{1} Load $M$ data elements of the two initial sorted arrays from the DRAM bank into the SPM buffers
\circlegf{2} Merge the two buffers into a third one, always placing the smaller data element across the two SPM buffers first.
\circlegf{3} Store the output buffer at the next position of the DRAM output. It is possible to use quicksort to sort the initial array chunks fitting into SPM, to speed up the mergesort execution.

\subsubsection{Hashing Algorithms on UPMEM}
\label{hashing}

In hashing, a hash function is applied to map each input element to a memory location in a hash table. An optimal hash function maps similar elements to different locations in the table. 
There is a vast choice of hash functions in the literature \cite{HashFunc77}, which are use-case dependent. 
To mitigate hashing collisions (i.e., when two elements of a hash table share the same hash value), a probing scheme is applied that determines the field to insert. When trying out \textit{linear probing}~\cite{LinearProbing} and \textit{cuckoo hashing}~\cite{Cuckoo} for this work, we found no meaningful difference in performance. Since linear probing is simpler to implement, we opt for using that in our evaluation.
Linear probing inserts the element into the next empty slot in the hash table. The drawback is that lookup time can grow with the occupancy of the hash table. Thus, we need to make sure it is sufficiently big. In practice we did not find this to be an issue for our use case.

The choice of hash functions poses a challenge, as we try to avoid costly operations like multiplication and division. A commonly used class of universal hash functions for example takes the form $f(x) = (ax \mod p) \mod m$ \cite{HashFunc77}, where $a, p$ are constants and $m$ is the table size. Since the performance of multiplication/division on PIM varies depending on operands \cite{Prim22}, it could theoretically be possible to find values $a, p$ with good performance. However, we found that fast performing values do not lead to good hash functions. Hash functions that rely on \textit{bit shifting}, \textit{addition} and \textit{xor} \cite{PearsonHash90}, are much better suited for our requirements.

Our goal here is again to minimize accesses to DRAM, while using as big transfers as possible. If we have one big hash table, it will not fit into SPM and we will have to fetch a small chunk for each access. To make it fit into SPM, we can partition it into smaller tables, by first partitioning the data into chunks. We also need to partition the data when distributing it. This means we have two applications for hashing in PIMDAL: (I) Insert the elements into a hash table, which we can probe when searching for them. (II) Partition the data into chunks, grouping equal elements together. 

\begin{figure}[h]
    \centering
    \includegraphics[width=0.9\linewidth]{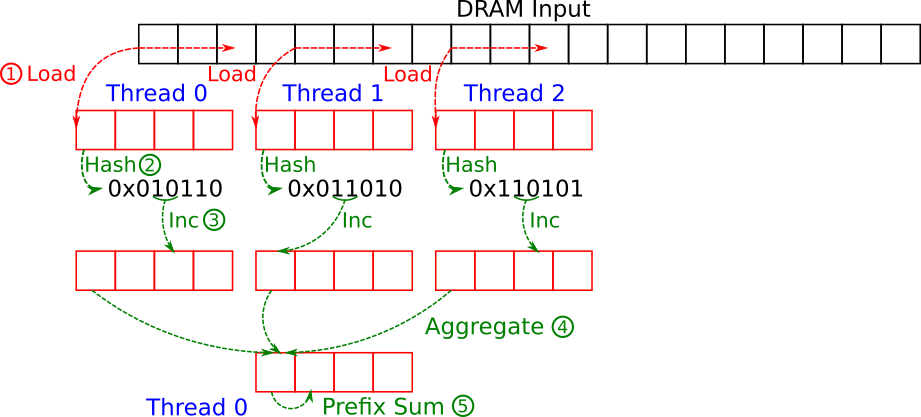}
    \caption{Multithreaded PIM implementation of hash partitioning step 1 and 2}
    \label{fig:join_part_1}
\end{figure}

\begin{figure}[h]
    \centering
    \includegraphics[width=0.9\linewidth]{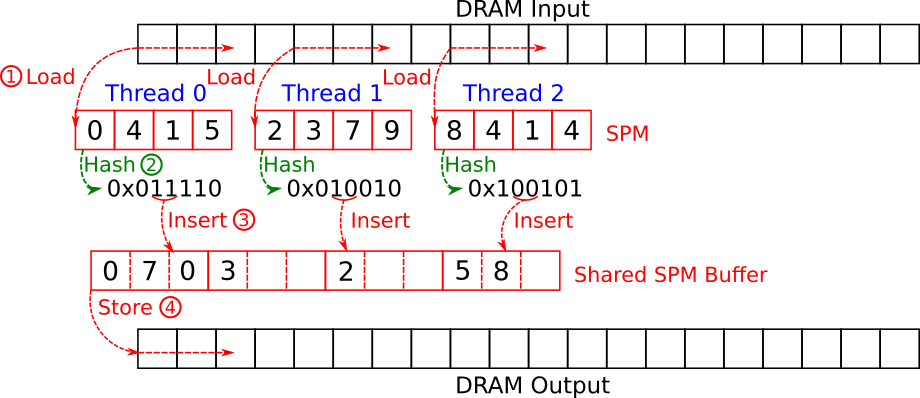}
    \caption{Multithreaded PIM implementation of hash partitioning step 3.}
    \label{fig:join_part_2}
    \vspace{-2.5mm}
\end{figure}

Hash partitioning consists of three main steps:
(1)~Calculating the size of each bucket, 
(2)~performing a prefix sum to find the offset in DRAM of each bucket and 
(3)~storing the elements at the bucket locations.
The first two steps, work as follows (Figure \ref{fig:join_part_1}):
\circlegf{1} Each thread loads $M$ data elements into an SPM buffer.
\circlegf{2} Each thread calculates the bucket of each element in the SPM buffer using a hash function.
\circlegf{3} Each thread uses this calculation to increment a local counter for each bucket size. 
\circlegf{4} The algorithm aggregates the local bucket size counters into a shared size and performs a prefix sum using a single thread on the shared counter to \circlegf{5} calculate the offset for each bucket.
The third step of hash partitioning, works as follows (Figure~\ref{fig:join_part_2}):
\circlegf{1}~Each thread loads the elements into a SPM buffer and \circlegf{2} calculates the bucket of each element in the buffer using the hash function.
\circlegf{3}~Each threads inserts its elements into the buckets in a shared SPM buffer.
\circlegf{4}~Once a bucket in the buffer is full, one thread writes its content back to the DRAM bank.

After the data has been partitioned, we can create hash tables that fit in the SPM for each partition. Each PIM-thread fetches a partition from DRAM, creates a hash table in the SPM and writes it back to the DRAM. In order to perform all the previously outlined steps, we have a memory requirement of $2\times$ the input size.

\subsection{Ordering Operator} \label{ordering}

The \texttt{ordering} operator is a straightforward application of our sorting algorithm. It takes an array in the PIM DRAM as an input and sorts it. Since TPC-H queries often require only returning a small subset of ordered data, single PIM-core sorting with \texttt{aggregation} on the host is often sufficient.

\subsection{Selection Operator}

\begin{figure}[h]
    \centering
    \includegraphics[width=0.8\linewidth]{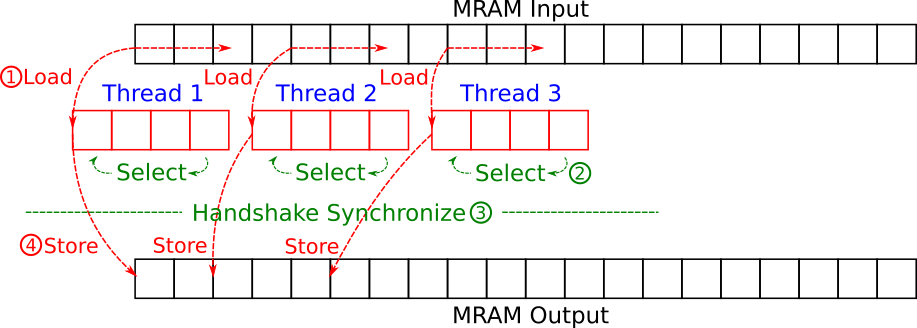}
    \caption{Multithreaded PIM implementation of \texttt{selection} operator.}
    \label{fig:select}
    \vspace{-2.5mm}
\end{figure}

The \texttt{selection} operator works by iterating through an array and filtering out the elements not satisfying a certain predicate, e.g. removing odd ones. It is usually used to select part of the rows for further processing. We base our implementation of the \texttt{selection} operator in \cite{Prim22} with a few optimizations. As shown in Figure~ \ref{fig:select}, the threads work on contiguous sections of the array. \circlegf{1}~Each thread loads $M$ elements into an SPM buffer. \circlegf{2}~It performs \texttt{selection} on the buffer, filtering out all elements not satisfying the predicate. \circlegf{3}~It waits for a handshake from the previous thread to get an offset in the output. Handshakes can be used to synchronize two threads \cite{UPMEMSynchronization}. One thread issues a \textit{notify}, while the other issues a \textit{wait\_for} instruction and they wait until both have been completed. This can be used to coordinate the exchange of data using shared memory. Each thread adds its own number of elements selected to the offset received and stores it in a variable for the next thread to receive. \circlegf{4}~It stores the filtered elements to the DRAM output array at the offset received.

\subsection{Aggregation Operator}

\texttt{Grouping} with \texttt{aggregation} groups rows with the same values in the specified key columns into a new output row. The remaining columns are aggregated based on an aggregation function. We implement four aggregation functions in our work:
(I)~\textbf{Unique} simply removes the duplicate rows;
(II)~\textbf{Count} counts the number of rows aggregated;
(III)~\textbf{Sum} sums up all the values in a column; and
(IV)~\textbf{Average} averages all the values of a column.
We will refer to \texttt{grouping} with \texttt{aggregation} simply as \texttt{aggregation} and show two implementations using sorting or hashing algorithms.

In the sorting-based \texttt{aggregation} method, the key columns are sorted in order to arrange duplicate elements contiguously, then the duplicate elements are filtered out. We first sort the elements using our Quicksort implementation (Section~\ref{sorting}). We then filter them the same way we do in the \texttt{selection} implementation (and similarly to the implementation in \cite{Prim22}). 
Instead of using a predicate, elements are now compared to subsequent ones. Compared to \texttt{selection}, the only difference is that intermediate results
need to be communicated between threads. Our implementation makes it possible to create user-defined aggregation functions.

\begin{figure}[h]
    \centering
    \includegraphics[width=0.9\linewidth]{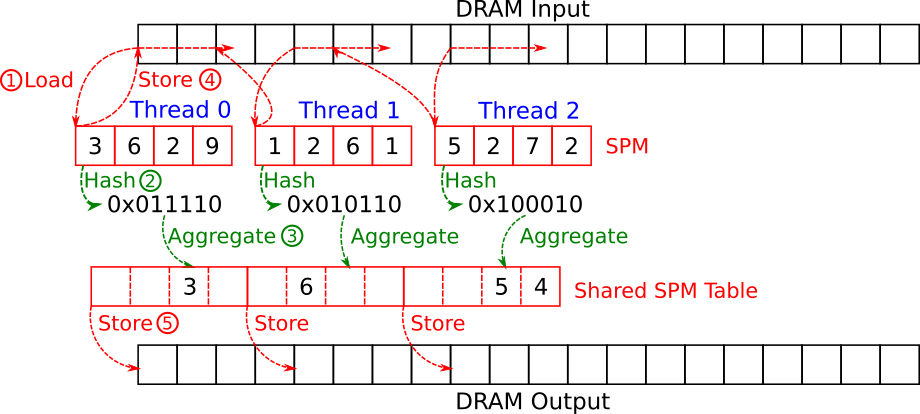}
    \caption{Multithreaded PIM implementation of hash \texttt{aggregation}.}
    \label{fig:hash_aggr}
\end{figure}

In the hash based \texttt{aggregation}  method, we insert all elements into a hash table, mapping duplicate elements to the same location.
Our implementation has three main steps (Figure~\ref{fig:hash_aggr}): 
\circlegf{1} Each thread loads the elements from the DRAM bank into an SPM buffer and \circlegf{2} calculates the hash table position of each element in the buffer.
\circlegf{3}~Each thread tries to insert the elements into a shared SPM hash table using mutexes for coordination: if the key of the entry is empty or the same as the input, the elements are aggregated; if an element cannot be inserted into the hash table due to conflicts, it remains in the buffer and is \circlegf{4}~written back to the input array.
\circlegf{5}~The aggregated columns in the SPM hash table are written to the output array in the DRAM bank.

Efficiently partitioning the hash table for \texttt{aggregation} is not straightforward \cite{muller2015cache}. The hash table size is equal to the output size, which is only known after completion. Naively partitioning the input can be unnecessary if the hash table already fits in the SPM. We ignore partitioning and show the impact in the evaluation.

\subsection{Join Operator}

\begin{figure}[h]
    \centering
    \includegraphics[width=0.9\linewidth]{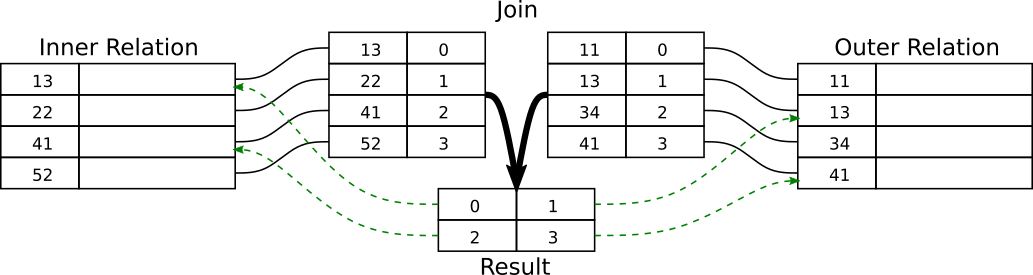}
    \caption{Result of joining two tables in PIMDAL.}
    \label{fig:join_method}
\end{figure}

The \texttt{join} operator takes tuples from multiple tables and concatenates them based on a specified condition. Inner \texttt{join} is the most common operation, where tuples are only returned if they have a matching key in both tables. Our \texttt{join} implementation produces the result shown in Figure \ref{fig:join_method}. First a key-value datatype is created with the key of each row and the value being a pointer to the row. Then, they are joined on the keys so that the result is now two pointers that describe how the rows of the two relations are joined.

In our work, we implement \textbf{hash} and \textbf{sort-merge joins} \cite{SortHash94, kim2009sort, muller2015cache}. Hash \texttt{join} first creates a hash table with the elements from the inner relation. This table is then probed with elements from the second relation. Elements with the same key get matched up this way. Sort-merge \texttt{join} first sorts the inner and outer relation to transform them into the same order. Then, it iterates through both relations, trying to match a tuple in the outer relation to one in the inner.

In our sort-merge \texttt{join} implementation, we first sort the two relations using multi PIM-core sorting. Each core then joins its part of the two relations that covers  the same keys:
(1) A thread loads $M$ elements of the inner relation into an SPM buffer $B_{inner}$. 
(2) It takes the first key and searches for an equal or greater key in the outer relation in DRAM. 
(3) It loads $M$ elements, starting from the key found, into a second SPM buffer $B_{outer}$. 
(4) It iterates through $B_{inner}$ and $B_{outer}$ searching for matching keys. For matches it writes the join indices to $B_{outer}$. 
(5) If $B_{outer}$ has been iterated through it writes the join indices stored there to the output DRAM array.

\begin{figure}[h]
    \centering
    \includegraphics[width=0.9\linewidth]{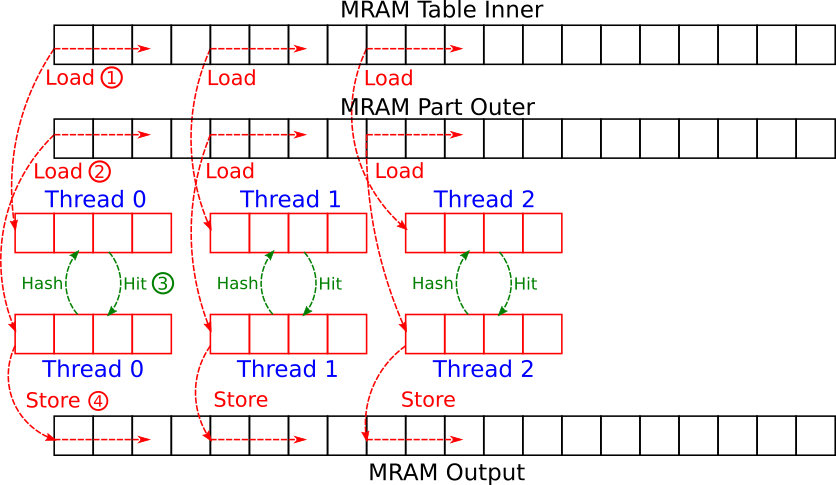}
    \caption{Multithreaded PIM implementation of hash \texttt{Join}.}
    \label{fig:join}
\end{figure}

For the hash \texttt{join} implementation we use the radix join algorithm \cite{manegold2002join}. This algorithm consists of two phases, the partitioning and the join phase. The partitioning has two functions: (I) Partitions are redistributed over all PIM-cores, so that each receives one bucket from each other core, corresponding to the same hash value. (II) Partition the hash tables so that they can fit into the SPM. Both the inner and outer relations have to be partitioned for this. Radix join partitions the data in multiple iterations, in order to improve performance. We explain the performance considerations for PIMDAL in the evaluation.
Hash join now consists of four steps:
(1) Partition the inner and outer relation and redistribute them over all PIM-cores.
(2) Partition the inner relation and insert each partition into a hash table as described in Section \ref{hashing}.
(3) Partition the outer relation so that we can use it to probe the hash table in the next step.
(4) Probe the partitioned hash tables with elements from the outer relation as follows (Figure~\ref{fig:join}): Each thread loads \circlegf{1} the partitioned hash tables and \circlegf{2} the partitioned data from the outer relation into SPM buffers. 
\circlegf{3} Each thread probes the hash table with an element from the outer partition. If there is a hit, join the elements to get the \texttt{join} indices.
\circlegf{4} Write the \texttt{join} indices back to the DRAM bank.

\section{Host Data Movement Optimization} \label{transfers}

To efficiently execute analytical queries, the communication between PIM-cores and to the host is crucial. We found this to be a key performance bottleneck when implementing some DB operations on UPMEM. In particular, naively performing data reordering and memory allocation on the host system leads to performance loss.

\subsection{Gather/Scatter Transfers}

\begin{figure}[h]
    \centering
    \includegraphics[width=\linewidth]{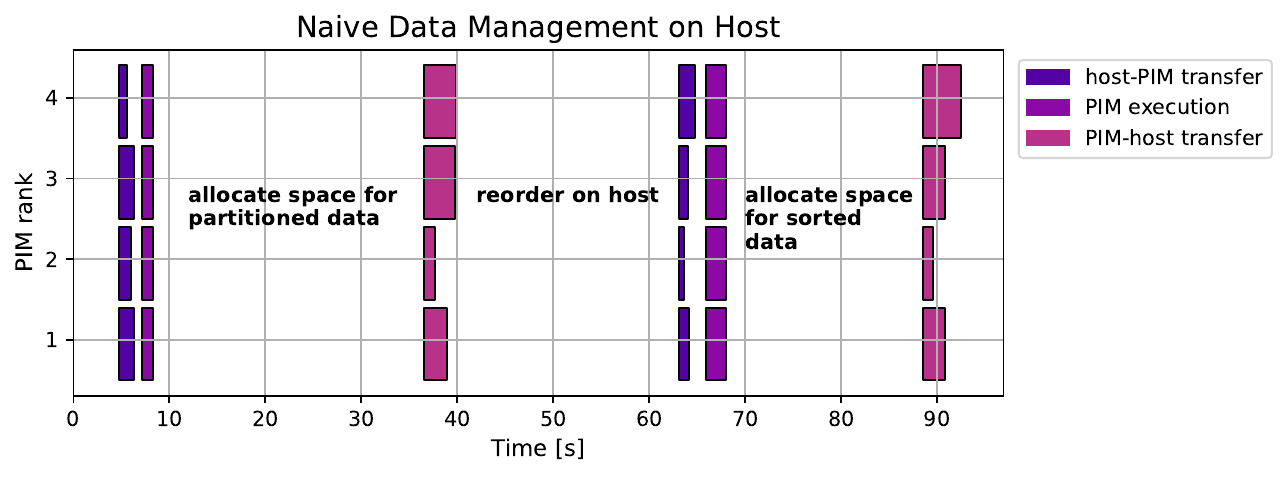}
    \caption{Example timeline of the standard data transfers between PIM memory and host.}
    \label{fig:naive_trans}
\end{figure}

For \texttt{ordering} and \texttt{join}, data has to be redistributed between PIM-cores during execution. The naive way to do this is by transferring the data from the PIM-cores to the host, reordering the data in the CPU memory, and transferring the reordered data back to the PIM-cores. Figure~\ref{fig:naive_trans} shows a timeline of transferring the data using this approach, with four host CPU threads copying data from and to the PIM ranks: The bars denote the time spent on data transfers between host and PIM system and on the PIM execution. The gaps between the bars denotes the idle time on the PIM-cores, waiting for the execution on the host CPU. We observe three major delays, creating a performance overhead. The delay at the beginning and end is caused by allocating the memory on the host for copying back data. The delay in the middle is caused by having to reorder data in the CPU memory for copying it back to the PIM system.

\begin{figure}[h]
    \centering
    \includegraphics[width=\linewidth]{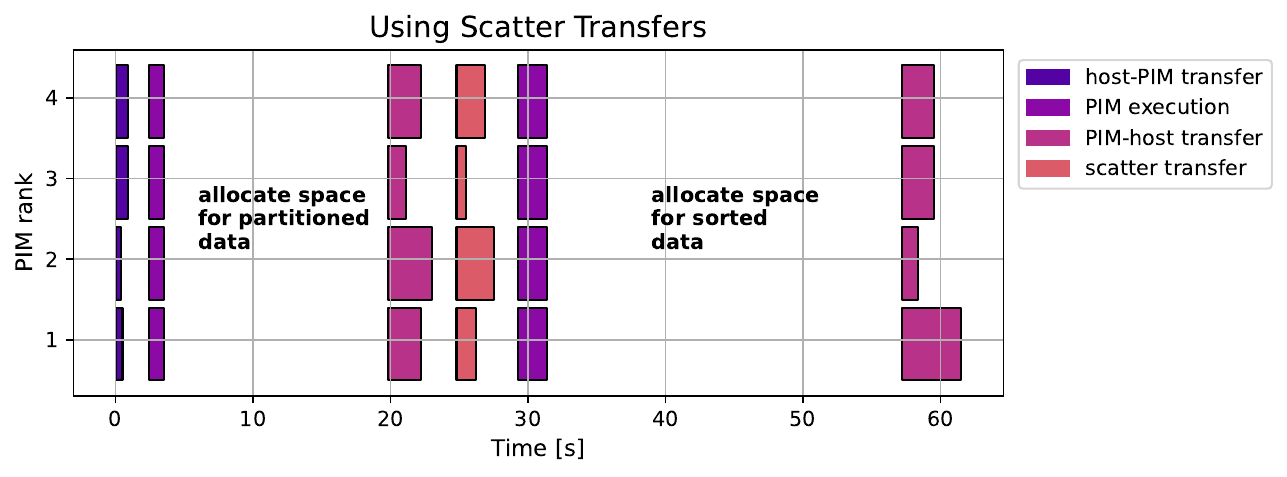}
    \caption{Example timeline after replacing the data reordering on the host with the scatter/gather transfer functionality.}
    \label{fig:scatter_trans}
\end{figure}

This reordering is a key bottleneck since it is restricted by the lower bandwidth of the host system. We make use of the new \textit{Scatter/Gather API} \cite{ScatterGather} in the UPMEM SDK to eliminate this overhead. This API enables precise copies of small, non-contiguous memory regions as opposed to the standard transfer methods. Previously, we had to arrange the partitioned fragment contiguously and copy them to the destination PIM-cores using parallel transfers. Now we can copy them directly to the destination PIM-cores. This improves the execution time as shown in Figure~\ref{fig:scatter_trans}. As it can be seen, the delay between the two transfers in the middle has been greatly shortened.

\subsection{Apache Arrow Memory Management} \label{mem_management}

\begin{figure}[h]
    \centering
    \includegraphics[width=\linewidth]{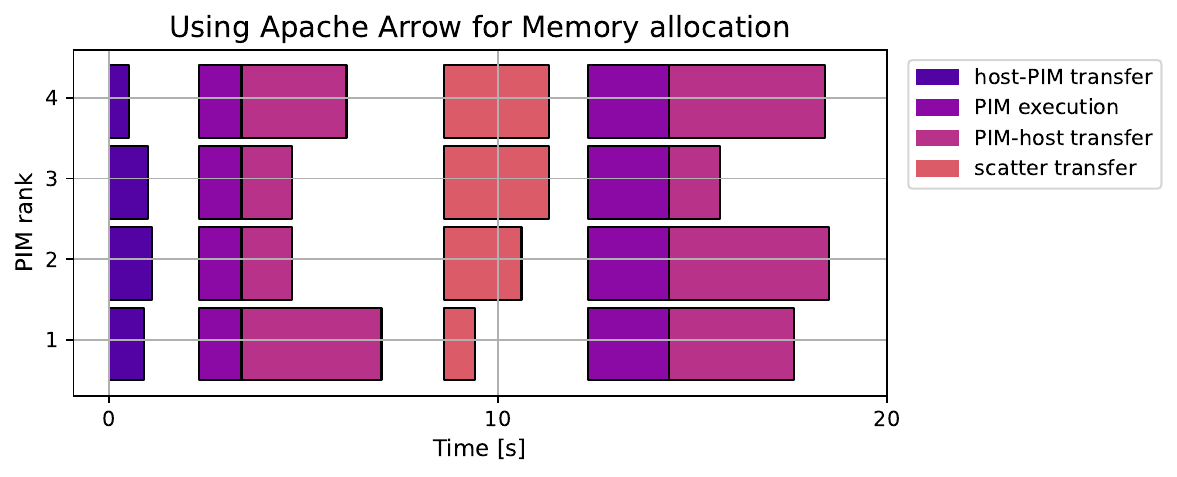}
    \caption{Example timeline of the improved execution time using Apache Arrow for host memory allocation.}
    \label{fig:arrow_trans}
\end{figure}

Even when using scatter transfers the performance is not ideal, due to the time needed for allocating space in the host memory, as seen in Figure \ref{fig:scatter_trans}. Although it halves the memory required for this step and thus the allocation overhead, this overhead is still considerable. The performance overhead is caused by the \textit{C++ standard library} containers and \textit{POSIX} \cite{Jemalloc06}. Especially with the big data sizes (order of 10 GB) we use in our implementation, it becomes unmanageable. To improve on this, we can use \textit{Apache Arrow}~\cite{ApacheArrow} to manage the memory allocation \cite{UPMEMOLAP}. It uses the \textit{jemalloc}~\cite{Jemalloc06} allocator, which uses a custom memory pool for allocation. This has less overhead and is considerably faster than the OS when using \textit{malloc} for larger memory sizes. We use \textit{Arrow} for allocating buffers that are transferred between the host and PIM system, significantly reducing allocation delay and
execution time, as can be seen in Figure \ref{fig:arrow_trans}.

\subsection{Asynchronous Transfers}

\begin{figure}[h]
    \centering
    \includegraphics[width=\linewidth]{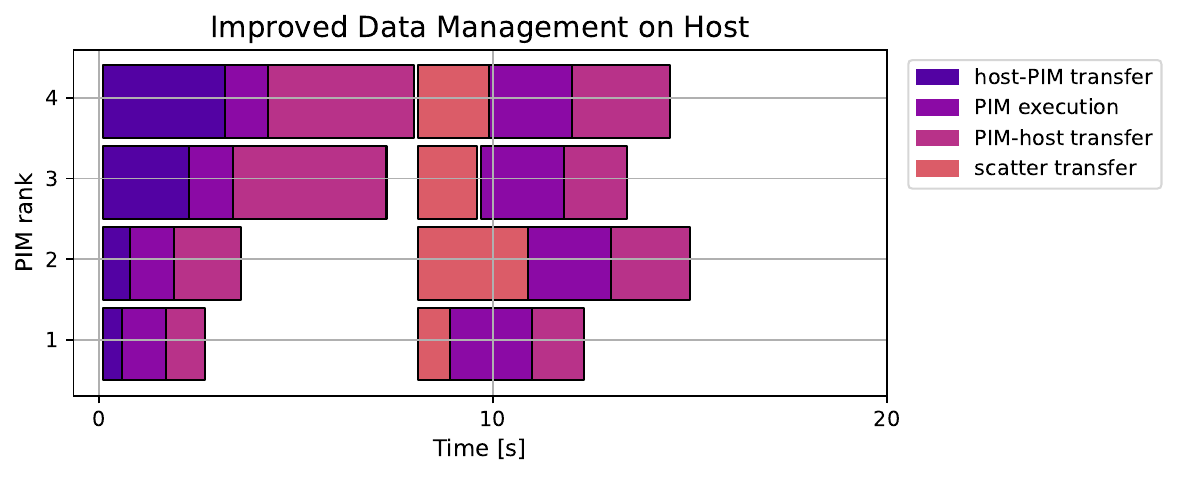}
    \caption{Example timeline showing the transfers and execution being performed asynchronously.}
    \label{fig:async_trans}
\end{figure}

Finally, we still see some overhead in Figure \ref{fig:arrow_trans}, due to the synchronous execution that blocks execution on some PIM ranks. Since some ranks receive data earlier, they could start executing without waiting for other ranks. This way the transfers back to the host are more spread out, which potentially utilizes the available bandwidth better. We can take advantage of the asynchronous transfers and execution supported by the UPMEM SDK. The results of all transfer optimizations can be observed in Figure \ref{fig:async_trans}. Compared to the synchronous execution we see how the bars for PIM execution and PIM-host transfer have different starting points. The shift in time of data transfers, with respect to each other, means better utilization of the bandwidth between PIM and host.

\section{Evaluation Methodology}

We first design a series of microbenchmarks to compare the different operator implementations in PIMDAL and identify the best design choices. Then we implement five TPC-H queries to compare our implementation to CPU and GPU reference implementations.

\subsection{Micro-Benchmarks}

In our micro-benchmarks we test the different operators separately. As a maximum size per PIM-core we chose 4M 32 bit integers for almost all operators, which corresponds to 16 MB of transferred data. The data moved around during operation is actually 32MB since we use key-value data types consisting of 2 integers. The only exception for this is the \texttt{join} operator, since hash \texttt{join} has a higher memory overhead. We use 4 MB of elements for the inner and 8 MB for the outer relation so that the total working size is 24 MB.

For the kernel execution we perform three measurements: (I) The kernel execution time for all operators. We use 2048 PIM-cores, and take the worst case execution time over all cores. (II) The IPC or occupancy of the PIM-core pipeline for all operators. The UPMEM SDK allows us to measure the number of instructions and cycles of code executed on the PIM system \cite{UPMEMSDK}. Dividing the number of instructions by the number of cycles gives us the IPC. If we use enough threads, as mentioned in Section \ref{arch}, the IPC should be an indicator for how much the pipeline is stalling due to inefficiencies of the operators. (III) A strong and a weak scaling analysis. For both we vary the number of PIM-cores and measure the overall execution time, including all data transfers. In the strong scaling analysis, we measure how execution improves when increasing the number of cores from 256 to 2048, while keeping the data size constant at 4GB. In the weak scaling analysis we measure how efficiently the workload can be scaled up, using 128 to 2048 PIM-cores and data sizes of 2 to 32 GB. The ratio of the execution times is the \emph{weak scaling efficiency}. We report a breakdown of the execution time of the individual steps. We also analyze how using asynchronous transfers and execution can improve execution time.

Finally we compare the performance of the DB operator implementation to a CPU and GPU reference. The key characteristics of all systems are shown in Table \ref{tab:system}.

\begin{table}[h]
    \centering
    \caption{Characteristics of the evaluated systems.}
    \resizebox{\linewidth}{!}{%
    \begin{tabular}{|c|c|c|c|}
        \hline
         & \textbf{\shortstack{UPMEM PIM \\ System}} & \textbf{\shortstack{Intel Xeon \\ Gold 6226R}} & \textbf{\shortstack{Nvidia \\ A6000 GPU}}\\
        \hline
        \hline
        \textbf{PEs} & 2048 DPUs & 16 (32 threads) & 108 SM\\
        \hline
        \textbf{Memory} & 131 GB & 128 GB & 48 GB\\
        \hline
        \textbf{Frequency} & 350 MHz & 2.9 GHz & 1.8 GHz\\
        \hline
        \textbf{Bandwidth} & 1.2 TB/s & 79 GB/s & 768 GB/s\\
        \hline
    \end{tabular}}
    \label{tab:system}
\end{table}

We implement the CPU reference for the DB operators from scratch for the micro-benchmarks, trying to follow the \textit{Apache Arrow} \cite{ApacheArrow} implementation. \texttt{Arrow} uses a data-centric approach to data analytics on processor-centric architectures. However, it is difficult to measure hardware performance metrics with it. We make sure our implementation performs similarly or better than \textit{Arrow}. For the GPU reference we use \textit{cuDF} \cite{cudf}. We compare once to both CPU and GPU using 8GB of data and once to CPU using 32GB of data. For \texttt{selection} we use a \emph{selectivity of $0.2$} and for \texttt{aggregation} \emph{50 groups}. These values are similar to what can be found in the TPC-H benchmark \cite{TPC}.

\subsection{TPC-H Benchmark}

The TPC-H benchmark \cite{TPC} is a decision support benchmark, consisting of business oriented ad-hoc queries. It is used for comparing commercial DBMS, covering practical use-cases.

We implement five queries from the TPC-H benchmark that consist of a mix of the implemented four DB operators. The selected queries do not rely on variable length datatypes. We generate the data for the queries using the TPC-H generator within DuckDB~\cite{DuckDB} with scaling factor 40, corresponding to \SI{40}{\giga\byte} total data size. It is then transformed into Apache Parquet format using the \textit{PyArrow} \cite{ApacheArrow}. Our PIM implementation loads the required data columns for each query onto the host using \textit{Apache Arrow}, before executing the query. The measured execution time is the time it takes to copy the data from the host to PIM memory, executing the query and copying the results back.

As a reference on CPU and GPU we implement the five queries using \textit{PyArrow}~\cite{PyArrow} and \textit{cuDF}~\cite{cudf}. These libraries also provide support for DBMSs with I/O from disk. In the CPU implementation we load the data from storage to RAM and then measure the execution time of the queries. For the GPU we also first load the data into the CPU RAM and copy it from there to the device. We report the execution time, including all loads to the GPU and back to the CPU. We do not use any query optimizer in any of the implementations. The goal is to isolate the performance induced by the underlying architecture, specifically the main memory bottleneck. DBMS developers use techniques like indexing to mitigate this issue. We expect PIM systems to profit similarly from this when optimized to the same degree in the future.

The TPC-H queries we implement are shown in Table \ref{tab:queries}. The key characteristics considered are the required DB operators and  arithmetic operations.

\begin{table}[h]
    \centering
    \caption{Number of occurrences per DB operator and whether addition and multiplications are used in each TPC-H query.}
    \begin{tabular*}{\linewidth}{|c@{\hspace{5pt}}@{\extracolsep{\fill}}|c|c|c|c|c|c|}
        \hline
        & \textbf{Sel} & \textbf{Aggr} & \textbf{Order} & \textbf{Join} & \textbf{Add} & \textbf{Mul} \\
        \hline
        \hline
        \textbf{Query 1} & $1\times$ & $8\times$ & & & $\bullet$ & $\bullet$ \\
        \hline
        \textbf{Query 3} & $3\times$ & $1\times$ & $1\times$ & $2\times$ & $\bullet$ & $\bullet$ \\
        \hline
        \textbf{Query 4} & $2\times$ & $2\times$ & & $1\times$ & $\bullet$ & \\
        \hline
        \textbf{Query 5} & $3\times$ & $1\times$ & & $5\times$ & $\bullet$ & $\bullet$ \\
        \hline
        \textbf{Query 6} & $3\times$ & $1\times$ & & & $\bullet$ & $\bullet$ \\
        \hline
    \end{tabular*}
    \label{tab:queries}
\end{table}

\section{Evaluation Results} \label{evaluation}

\subsection{Micro-Benchmarks} \label{microbench}

\begin{figure}[h]
\centering
\includegraphics[width=1\linewidth]{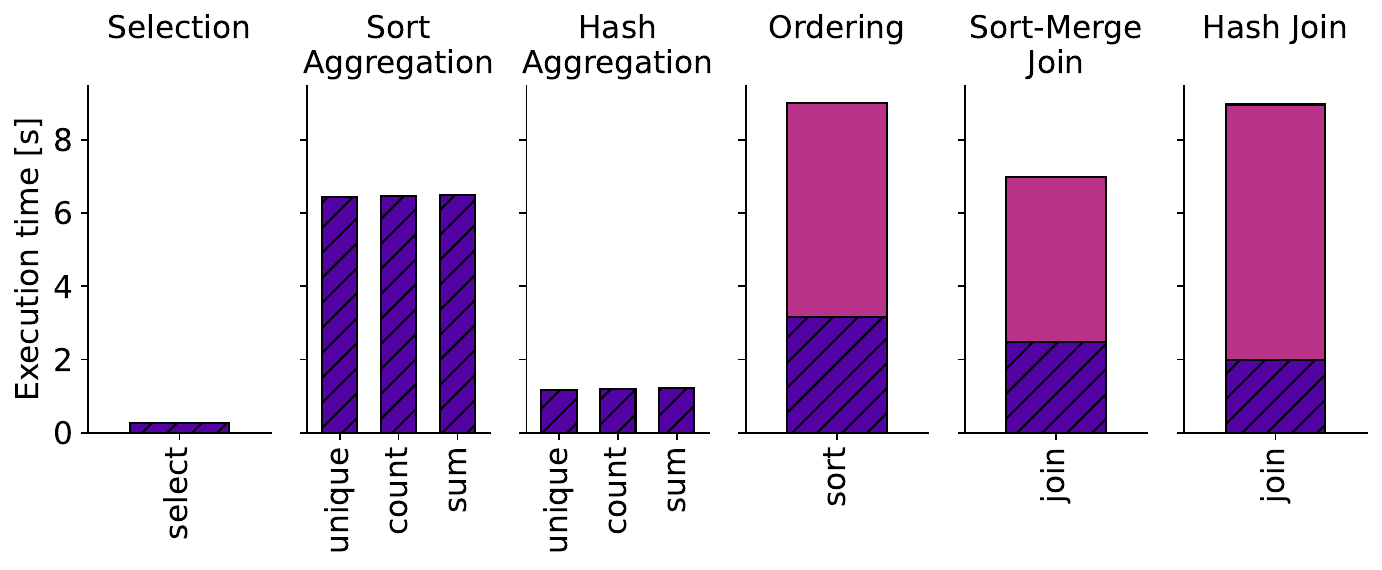}
\caption{Kernel execution time comparison. For \texttt{ordering} and both \texttt{join} implementations the two phases of execution are shown together.}
\label{fig:kernels}
\end{figure}

In our DB operator evaluation we first focus on the single PIM-core performance and then look at the whole system. Figure \ref{fig:kernels} shows the kernel execution time for the different DB operators. For the operators requiring two steps of execution (partitioning and final operation), it shows the time taken by both steps.

\begin{figure}[h]
\centering
\includegraphics[width=\linewidth]{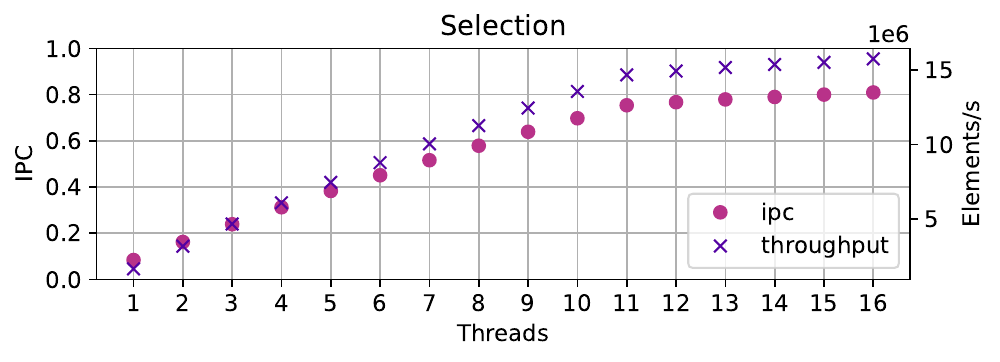}
\caption{IPC and throughput scaling of selection for a varying number of threads.}
\label{fig:ipc_sel}
\end{figure}

Figure \ref{fig:ipc_sel} shows an in-depth view of the performance (right y-axis) and IPC (left y-axis) of \texttt{selection} with increasing number of threads. The performance increases linearly until 11 threads, which is the minimum number required for full pipeline occupancy. After that point, the increase slows down considerably. The increase in performance corresponds to the increase in IPC. The main source of stalls in selection are the memory transfers. These are effectively mitigated by increasing the number of threads, improving the usage of compute resources.

When comparing sort and hash \texttt{aggregation}, we see the implementation using hashing being about $5 \times$ faster. The overhead of the aggregation functions is negligible overhead.

\begin{figure}[h]
\centering
\includegraphics[width=\linewidth]{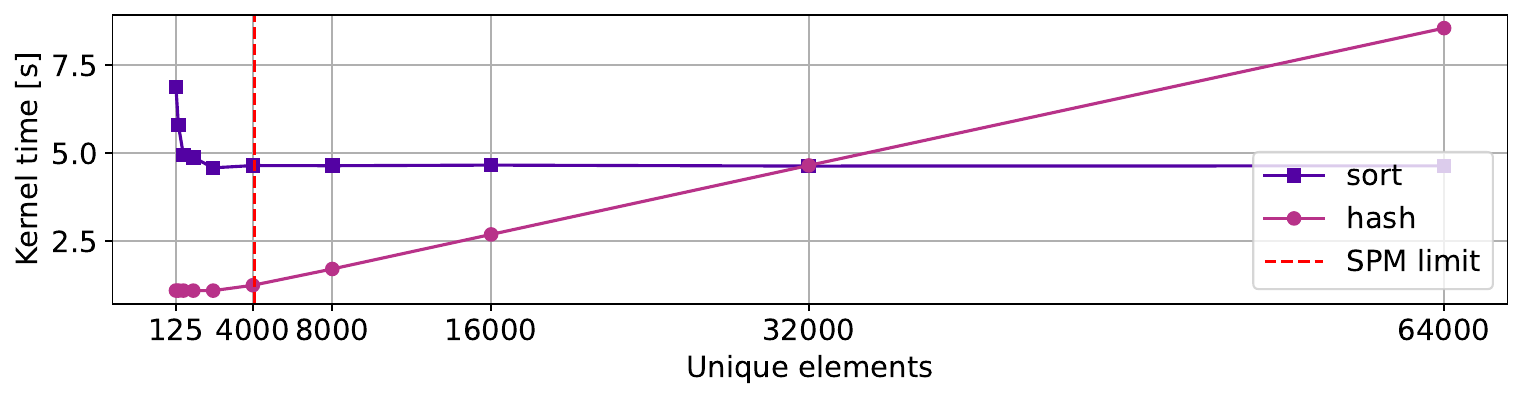}
\caption{Comparison of \texttt{aggregation} execution time using hashing and sorting for different numbers of unique input elements.}
\label{fig:aggr_comp}
\end{figure}

Figure \ref{fig:aggr_comp} shows the execution time (y-axis) for hash and sort based \texttt{aggregation} for a varying number of unique elements (x-axis). More unique elements mean more groups or bigger output size. As we can see, the execution time of hash \texttt{aggregation} increases with increasing size, while it remains constant for the sort based one. This is because the sorting algorithm partitions the data, to improve SPM usage. However, this partitioning is inefficient for a low number of unique elements, which is why the execution time is high. The hash based algorithm simply repeats all the steps if the data cannot fully fit into the SPM. To optimize performance of aggregation we could switch between the two implementations. Alternatively, we can partition the data for hash aggregation, as done on conventional systems to similarly improve cache efficiency \cite{muller2015cache}. However, this requires the use of a query optimizer, to analyze the data and chose the best performing algorithm. The need for efficient SPM usage is similar to conventional systems and suggests that we can reuse current optimizers for PIM.

\takeaway{SPM or cache utilization still plays a role in the performance of PIM architectures, as the example of hash and sort based \texttt{aggregation} shows.}

\begin{figure}[h]
\centering
\includegraphics[width=\linewidth]{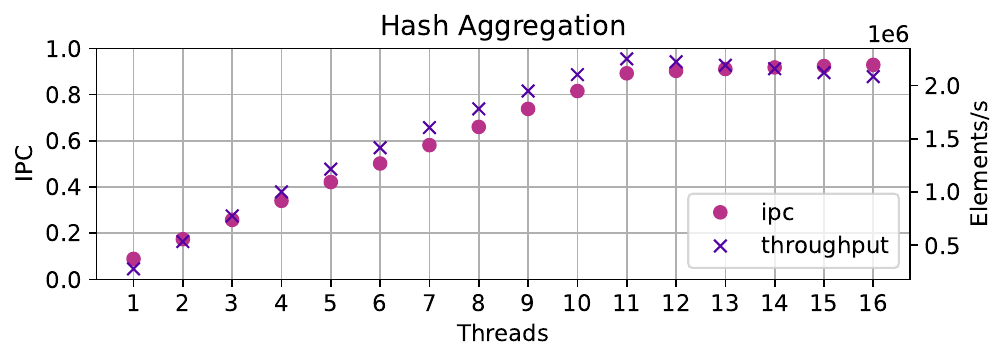}
\caption{IPC and throughput scaling of hash \texttt{aggregation} for a varying number of threads.}
\label{fig:ipc_haggr}
\end{figure}

The performance of hash \texttt{aggregation} improves similarly to \texttt{selection} with increasing number of threads (Figure \ref{fig:ipc_haggr}). However, after 11 threads the performance starts decreasing. The IPC on the other hand keeps increasing. This suggest that contention for the mutexes, needed for synchronization, erases the gains from increasing the thread count.

\begin{figure}[h]
\centering
\includegraphics[width=\linewidth]{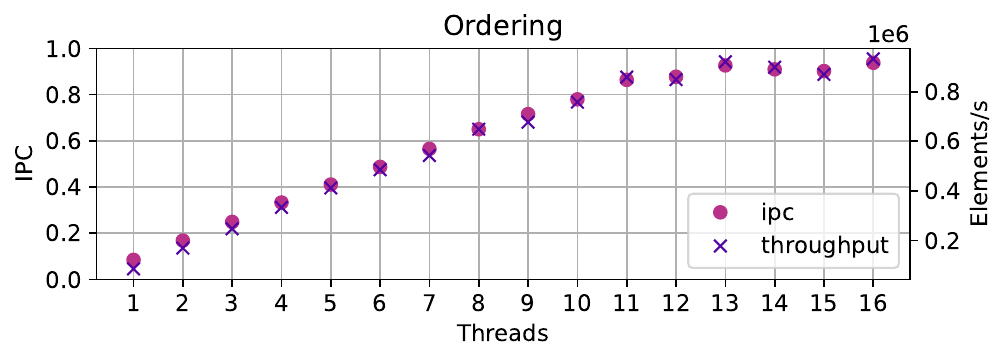}
\caption{IPC and throughput scaling of ordering for a varying number of threads.}
\label{fig:ipc_ordering}
\end{figure}

The performance of sort \texttt{aggregation} is basically equal to just sorting. Looking at Figure \ref{fig:ipc_ordering}, we see that the performance mirrors the IPC, which reaches up to 0.92. This suggests that sorting is a perfect fit for PIM architectures. The ratio of data accesses to computation is ideal for the characteristics of PIM.

When comparing the performance of iterative to recursive quicksort, we found the iterative implementation to perform about $1.1 \times$ better. The implementation is also simpler, since we do not have to increase the stack size. Increasing the buffer size from 64 to 256, we observe a speedup of $1.1 \times$. Bigger transfer sizes increase performance of sorting, but not significantly.

\begin{figure}[h]
\centering
\includegraphics[width=\linewidth]{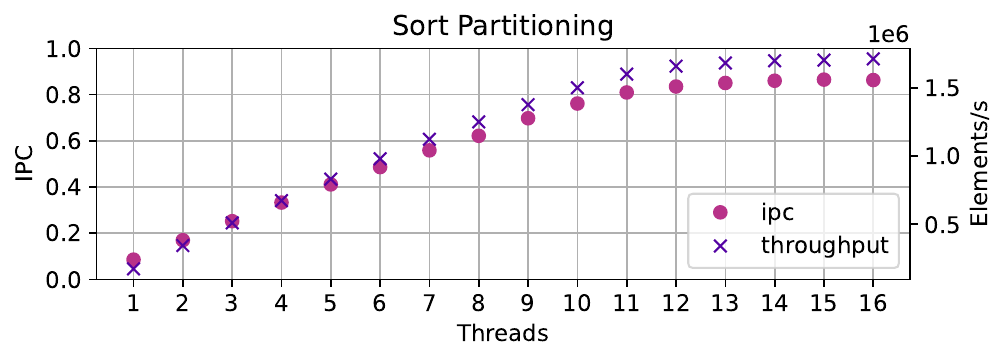}
\caption{IPC and throughput scaling of sort partitioning for a varying number of threads.}
\label{fig:ipc_spart}
\end{figure}

Comparing sort-merge \texttt{join} with hash \texttt{join}, we observe sort-merge \texttt{join} being slightly faster. This differs from the behavior observed on conventional architectures bottlenecked by bandwidth, for the two algorithms \cite{SortHash94, kim2009sort, muller2015cache}. The first reason for this is the change in the ratio of bandwidth to arithmetic performance. Hashing on PIM is costly, while the additional memory accesses in sorting are cheaper. The second reason is the limited SPM. This can be observed especially when looking at hash compared to sort partitioning. The IPC of our sort partitioning implementation increases uniformly with the number of threads (Figure \ref{fig:ipc_spart}). Sorting allows each thread to have its own resources, at the cost of higher algorithmic complexity.

\begin{figure}[h]
\centering
\includegraphics[width=\linewidth]{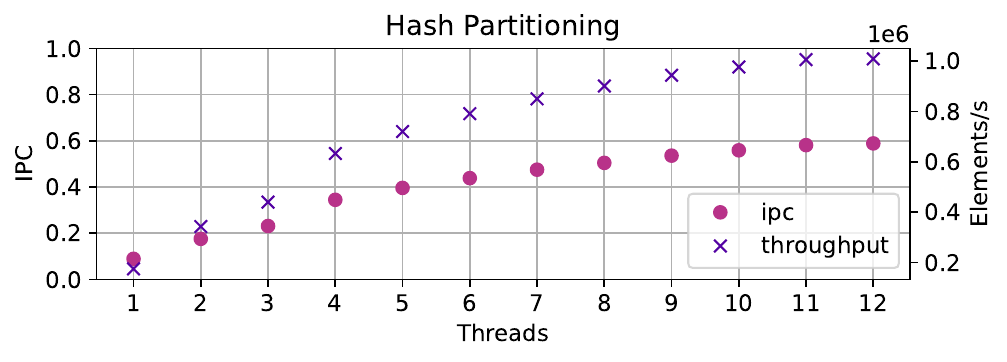}
\caption{IPC and throughput scaling of hash partitioning for a varying number of threads.}
\label{fig:ipc_hpart}
\end{figure}

Our hash partitioning implementation is less efficient, as seen from the IPC (Figure \ref{fig:ipc_hpart}). It uses the multi-iteration radix-cluster algorithm \cite{manegold2002join}, which in each iteration assigns data into different buckets. Using more buckets requires less iterations, however, the bucket size also shrinks. While radix-cluster improves TLB misses on conventional CPUs, the issue on PIM is the memory access time behavior. As we mention in Section \ref{arch}, too small data transfers can reduce the performance on PIM. If we use a many buckets, the cache and transfer size is very small, due to the limited SPM. This creates a trade-off with the number of iterations. Since the SPM is shared between threads, we create a shared bucket cache. However, this requires synchronization, which is the reason for the low observed IPC. When sweeping the number of buckets per iteration from 8 to 64, we find 32 to be the optimal number. While PIM characteristics are different, our analysis suggests that prior work on data partitioning \cite{muller2015cache} is still relevant.

\takeaway{The most efficient algorithms for PIM can differ from the state-of-the-art on other architectures. One such example is sort-merge compared to hash join.}

\observation{The performance of all operators improves with the number of threads for all operators for up to 11 threads. For more than 11 threads, gains are often insignificant or negative.}

\begin{figure*}[ht]

\begin{subfigure}{\linewidth}
\includegraphics[width=0.8\linewidth]{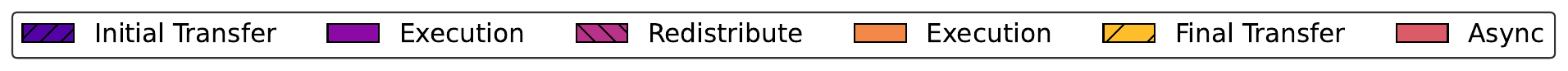}
\end{subfigure}

\begin{subfigure}{0.33\linewidth}
\centering
\includegraphics[width=\linewidth]{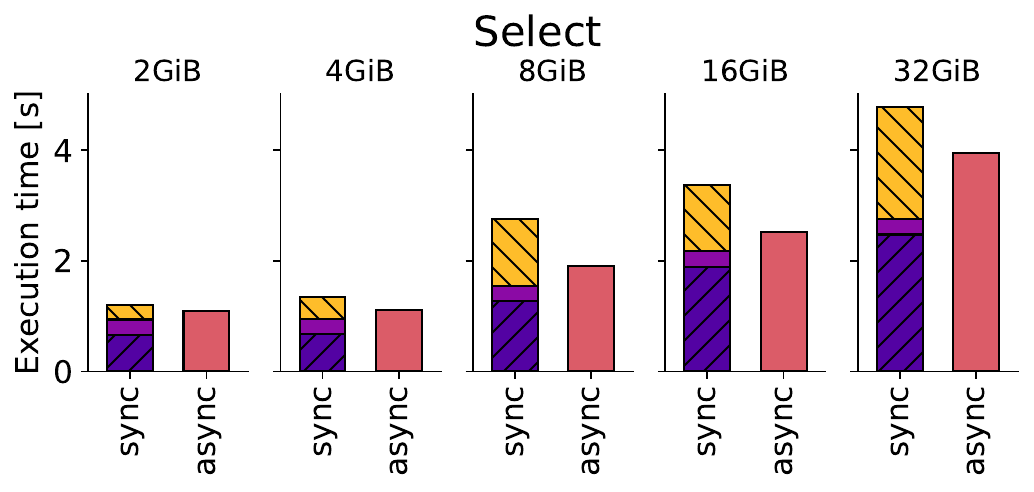}
\end{subfigure}
\begin{subfigure}{0.33\linewidth}
\centering
\includegraphics[width=\linewidth]{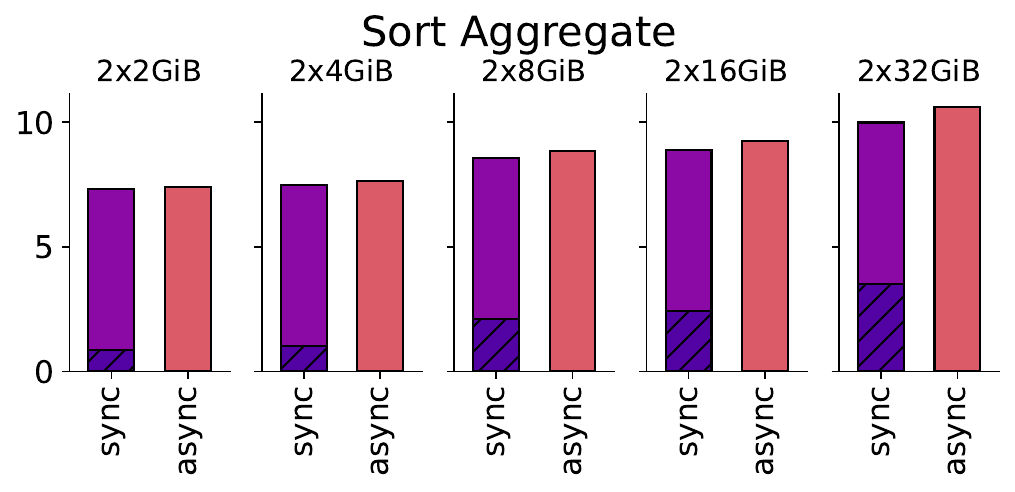}
\end{subfigure}
\begin{subfigure}{0.33\linewidth}
\centering
\includegraphics[width=\linewidth]{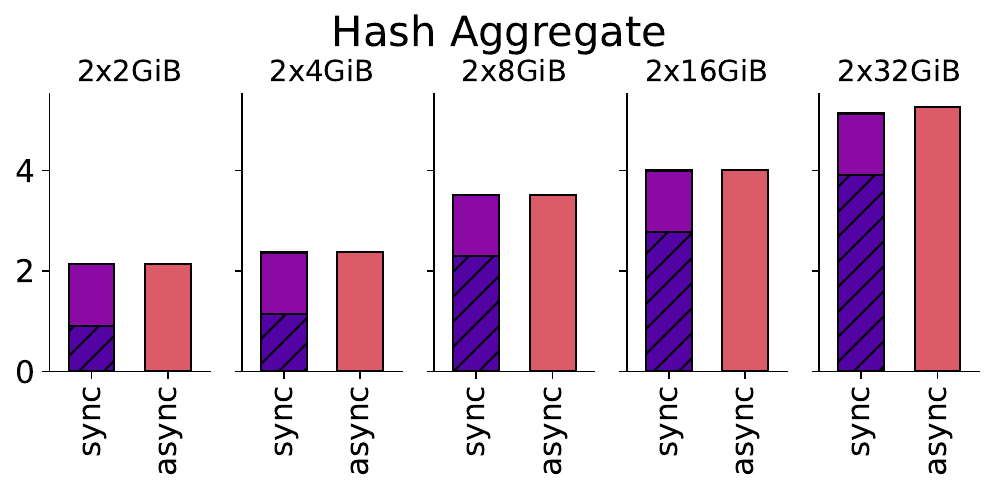}
\end{subfigure}

\begin{subfigure}{0.33\linewidth}
\centering
\includegraphics[width=\linewidth]{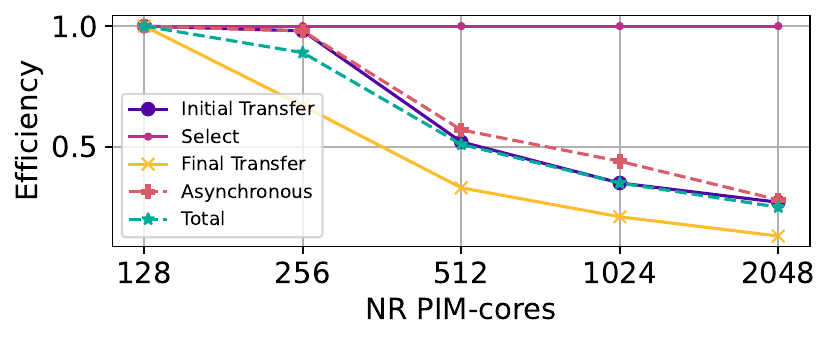}
\caption{Selection operator}
\label{fig:sel_weak}
\end{subfigure}
\begin{subfigure}{0.33\linewidth}
\centering
\includegraphics[width=\linewidth]{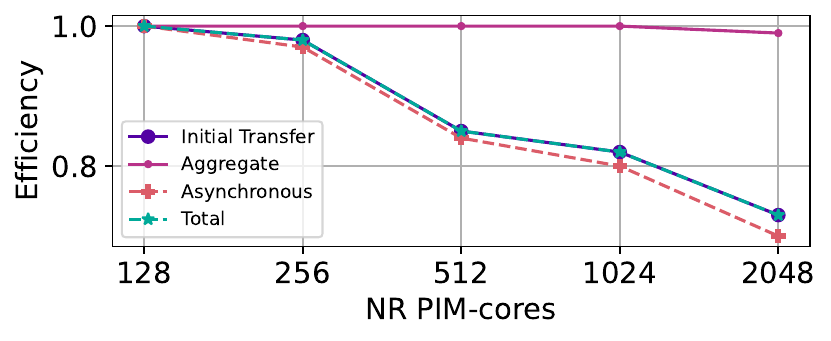}
\caption{Sort aggregation operator}
\label{fig:saggr_weak}
\end{subfigure}
\begin{subfigure}{0.33\linewidth}
\centering
\includegraphics[width=\linewidth]{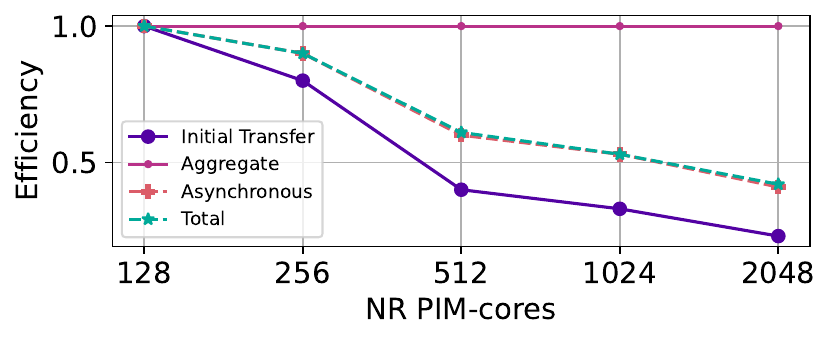}
\caption{Hash aggregation operator}
\label{fig:haggr_weak}
\end{subfigure}

\begin{subfigure}{0.33\linewidth}
\centering
\includegraphics[width=\linewidth]{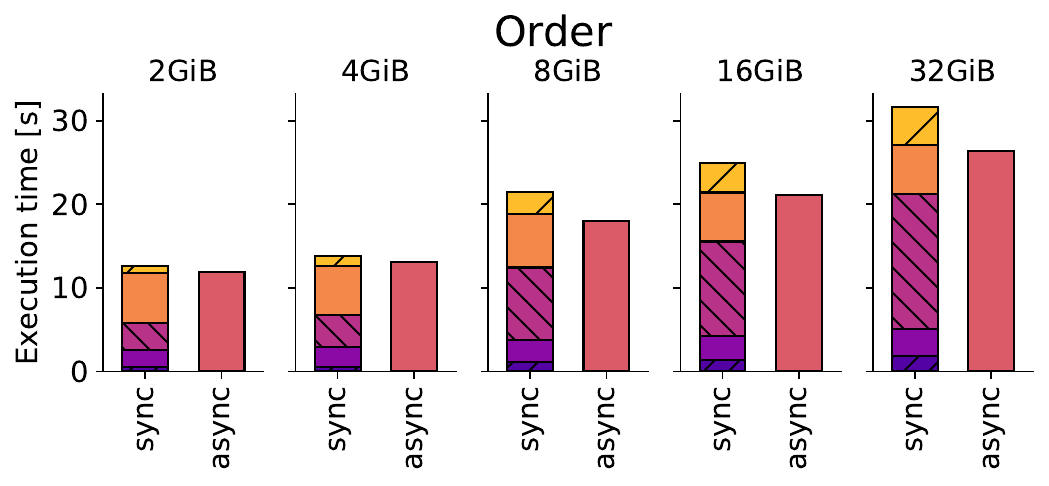}
\end{subfigure}
\begin{subfigure}{0.33\linewidth}
\centering
\includegraphics[width=\linewidth]{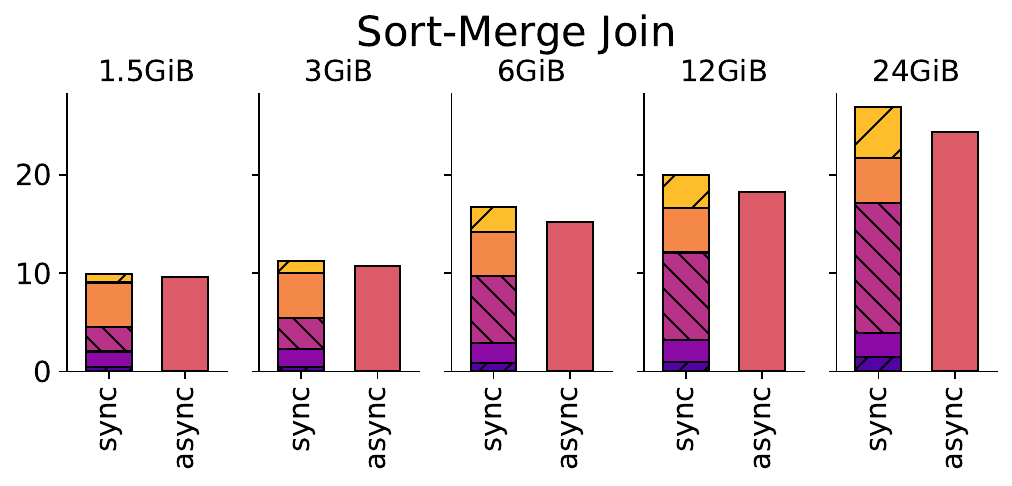}
\end{subfigure}
\begin{subfigure}{0.33\linewidth}
\centering
\includegraphics[width=\linewidth]{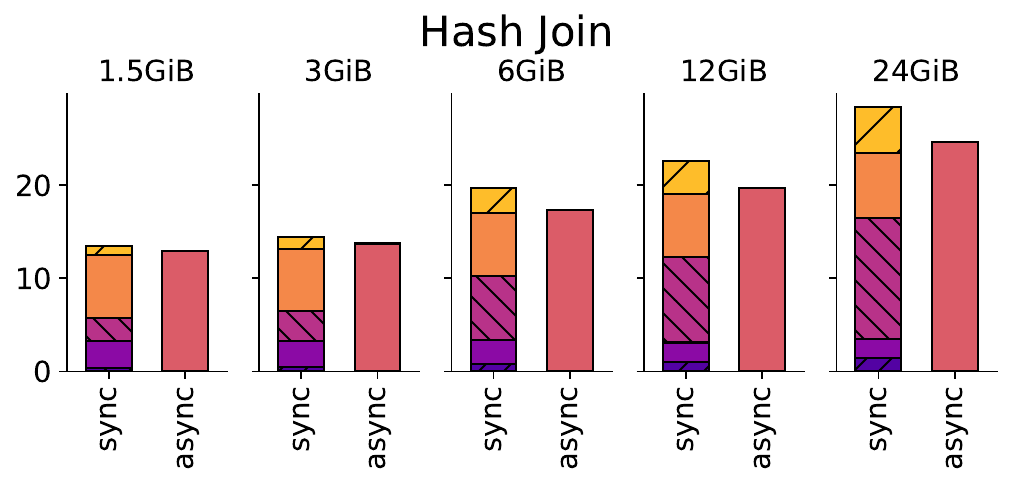}
\end{subfigure}

\begin{subfigure}{0.33\linewidth}
\centering
\includegraphics[width=\linewidth]{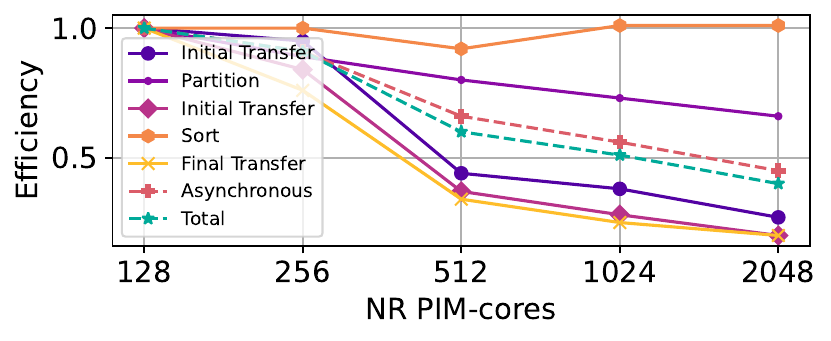}
\caption{Ordering operator}
\label{fig:order_weak}
\end{subfigure}
\begin{subfigure}{0.33\linewidth}
\centering
\includegraphics[width=\linewidth]{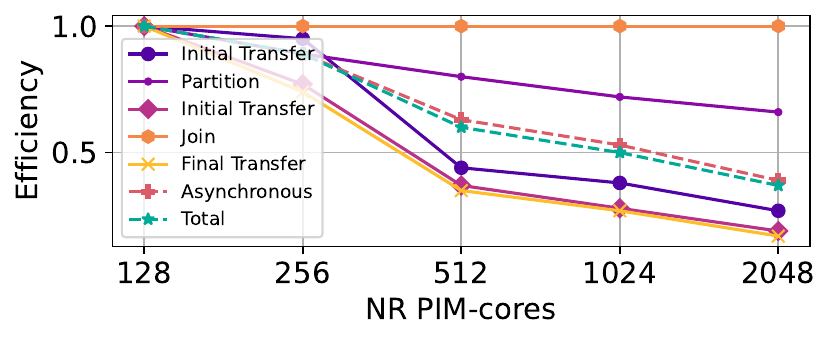}
\caption{Sort-merge join operator}
\label{fig:smjoin_weak}
\end{subfigure}
\begin{subfigure}{0.33\linewidth}
\centering
\includegraphics[width=\linewidth]{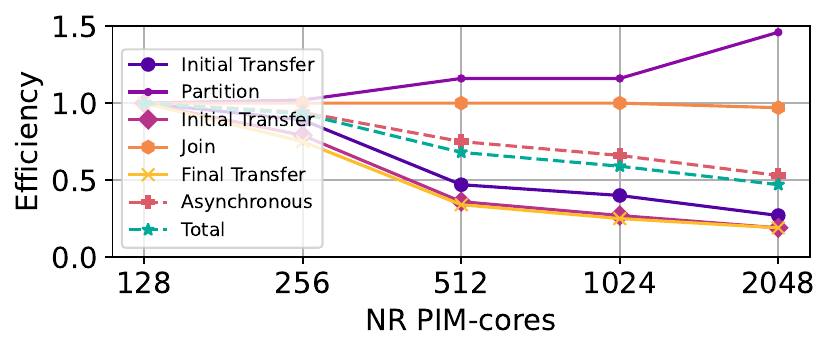}
\caption{Hash join operator}
\label{fig:hjoin_weak}
\end{subfigure}

\caption{Weak scaling comparison of operator implementations using execution time (bar plots) and efficiency (line plots).}
\label{fig:weak_scaling}
\end{figure*}

Figure \ref{fig:weak_scaling} shows the weak scaling analysis (bar plots) and the efficiency (line plots), broken down into execution and transfer times. The efficiency, which shows how well one can increase the problem size with additional resources, usually decreases due to increased overheads. This means it is usually between 0 and 1. For \texttt{selection} we see that the main bottleneck are the data transfers with the host (\ref{fig:sel_weak}). When increasing the number of PIM cores we increase the number of DIMMs, which increases the bandwidth. However, from the execution time we see that this increase is not proportional.

In hash \texttt{aggregation} the transfer time is also a key bottleneck in the execution time (\ref{fig:haggr_weak}). This is not the case for sort \texttt{aggregation} because of the increased kernel execution time (\ref{fig:saggr_weak}) compared to hash \texttt{aggregation}. However, we see that the transfers to the host are negligible, since the data has been reduced considerably. This is also the reason \texttt{aggregation} cannot profit from asynchronous transfers. We see that it only improves execution time when there are multiple, bigger transfers, that can be offset in time.

For \textit{ordering} the key bottleneck is the data redistribution between PIM-cores (\ref{fig:order_weak}). As we show in Section \ref{transfers}, optimizing this step is crucial for the overall performance. If we ignore the redistribution step, we see that the execution times exceed the transfer times. This shows that the operator is significantly more complex than the previous ones.

The results for sort-merge (\ref{fig:smjoin_weak}) and hash \texttt{join} (\ref{fig:hjoin_weak}) look similar to \texttt{ordering}. This is not surprising since sort-merge \texttt{join} is very similar to \texttt{ordering}. Hash \texttt{join} only differs in terms of PIM execution, as we previously outlined. We see that all these operators are limited by data redistribution between PIM-cores in current systems.

\recommendation{Performing data transfers and execution asynchronously improves performance when data has to be transferred \textbf{both} from and to the host.}

\begin{figure*}[ht]

\begin{subfigure}{\linewidth}
\includegraphics[width=0.8\linewidth]{figures/legend.pdf}
\end{subfigure}

\begin{subfigure}{0.33\linewidth}
\centering
\includegraphics[width=\linewidth]{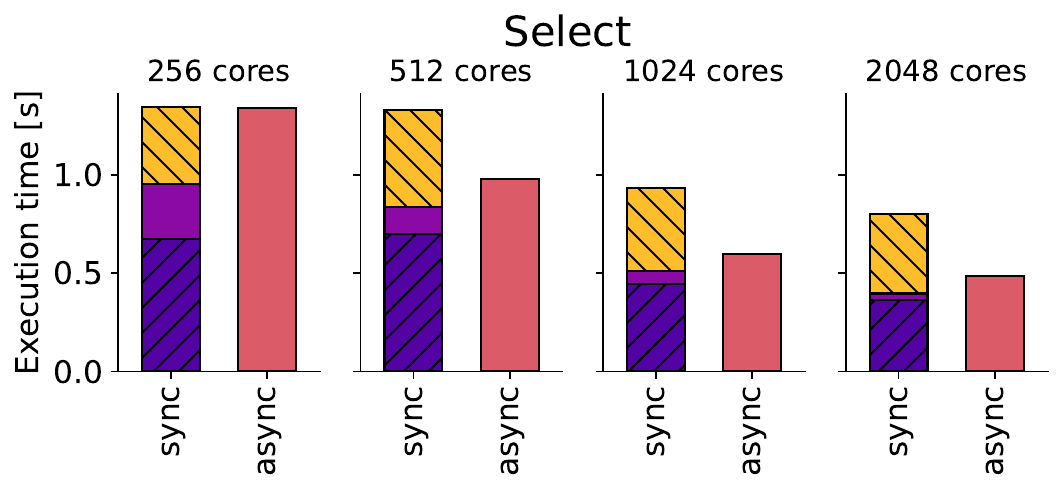}
\end{subfigure}
\begin{subfigure}{0.33\linewidth}
\centering
\includegraphics[width=\linewidth]{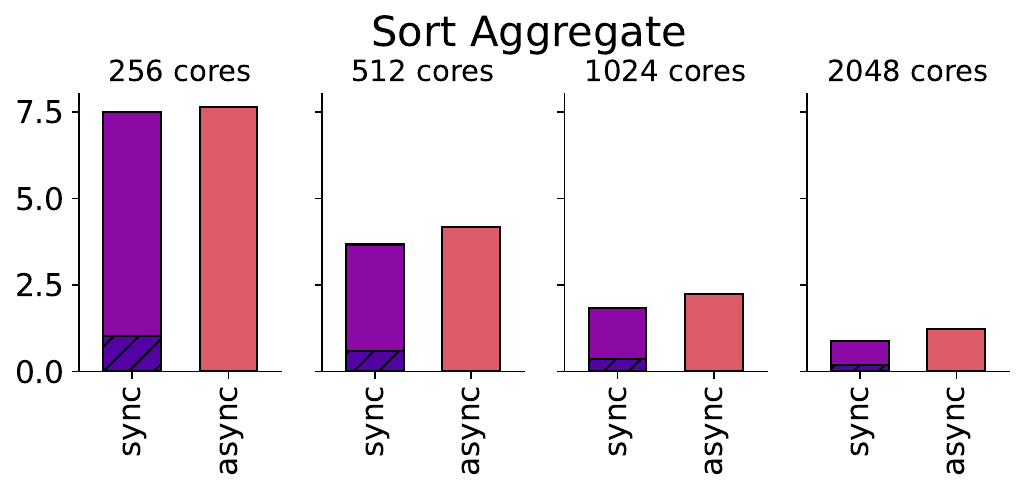}
\end{subfigure}
\begin{subfigure}{0.33\linewidth}
\centering
\includegraphics[width=\linewidth]{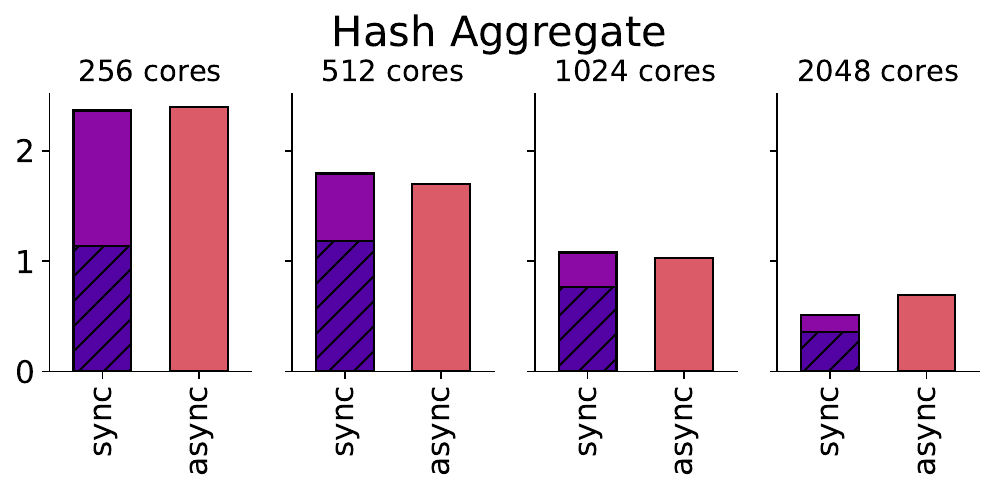}
\end{subfigure}

\begin{subfigure}{0.33\linewidth}
\centering
\includegraphics[width=\linewidth]{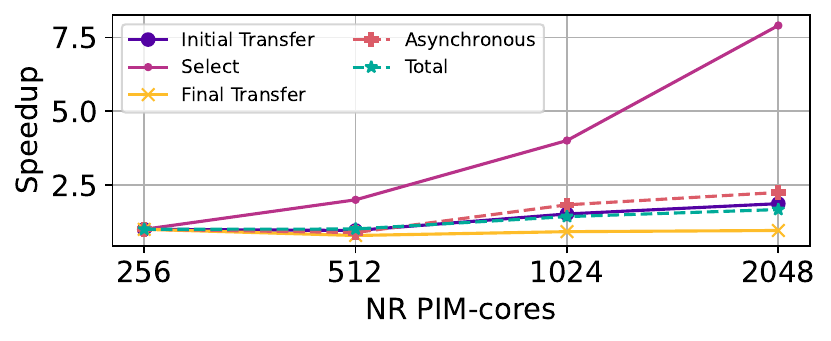}
\caption{Selection operator}
\label{fig:sel_strong}
\end{subfigure}
\begin{subfigure}{0.33\linewidth}
\centering
\includegraphics[width=\linewidth]{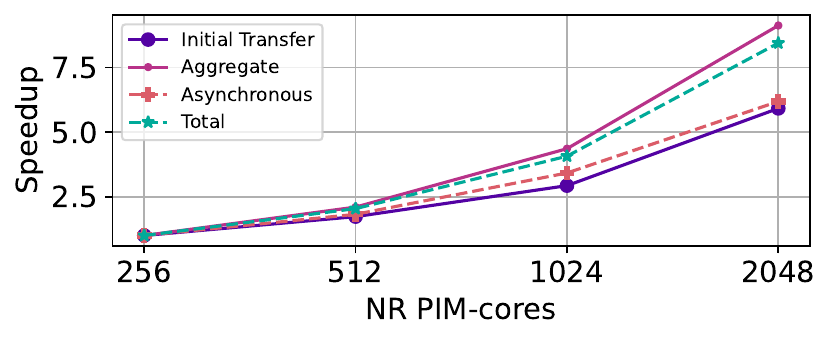}
\caption{Sort aggregation operator}
\label{fig:saggr_strong}
\end{subfigure}
\begin{subfigure}{0.33\linewidth}
\centering
\includegraphics[width=\linewidth]{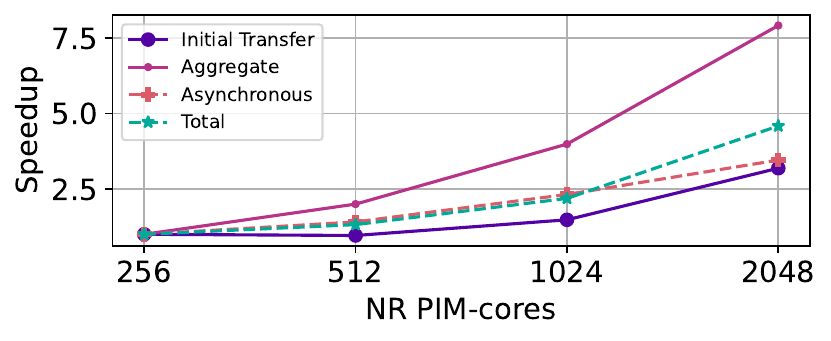}
\caption{Hash aggregation operator}
\label{fig:haggr_strong}
\end{subfigure}

\begin{subfigure}{0.33\linewidth}
\centering
\includegraphics[width=\linewidth]{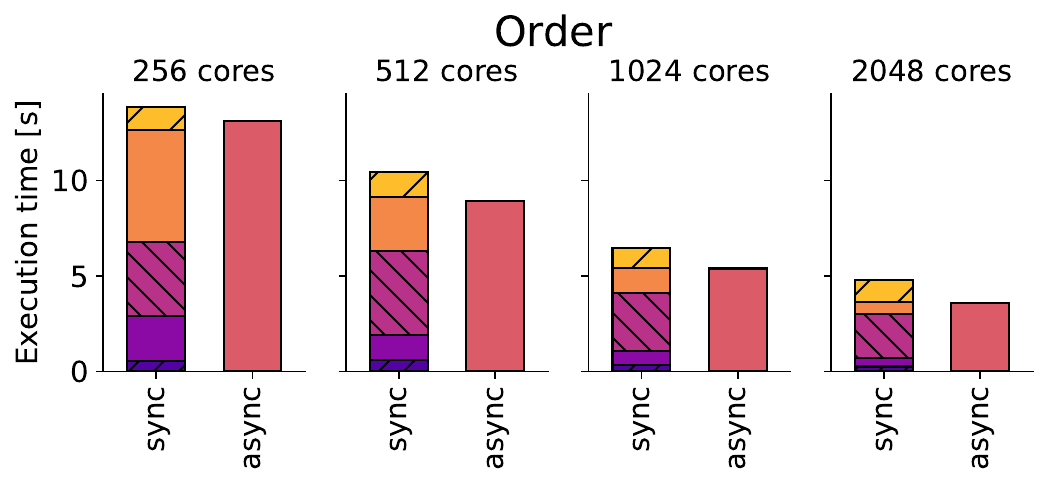}
\end{subfigure}
\begin{subfigure}{0.33\linewidth}
\centering
\includegraphics[width=\linewidth]{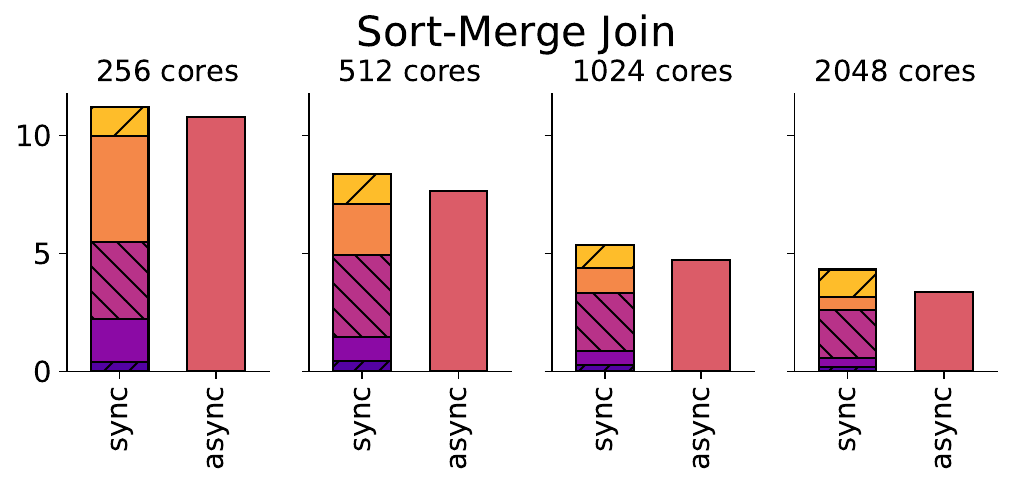}
\end{subfigure}
\begin{subfigure}{0.33\linewidth}
\centering
\includegraphics[width=\linewidth]{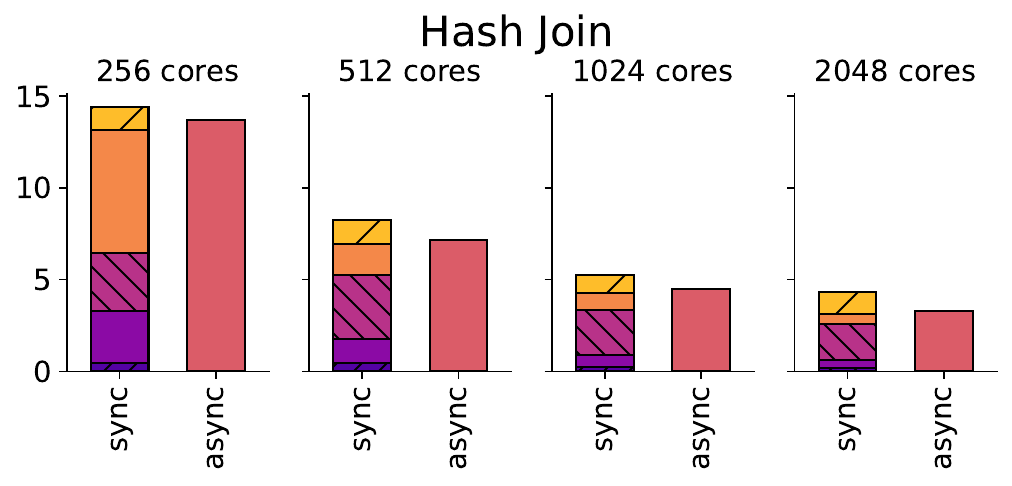}
\end{subfigure}

\begin{subfigure}{0.33\linewidth}
\centering
\includegraphics[width=\linewidth]{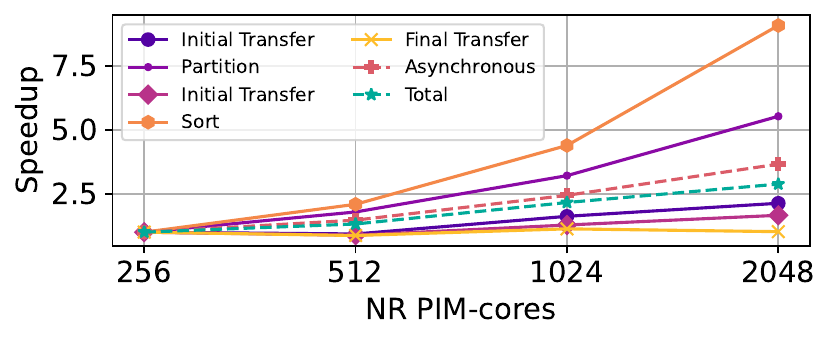}
\caption{Ordering operator}
\label{fig:order_strong}
\end{subfigure}
\begin{subfigure}{0.33\linewidth}
\centering
\includegraphics[width=\linewidth]{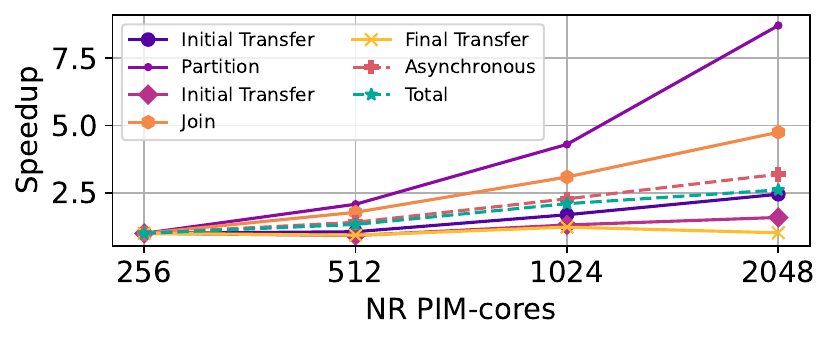}
\caption{Sort-merge join operator}
\label{fig:smjoin_strong}
\end{subfigure}
\begin{subfigure}{0.33\linewidth}
\centering
\includegraphics[width=\linewidth]{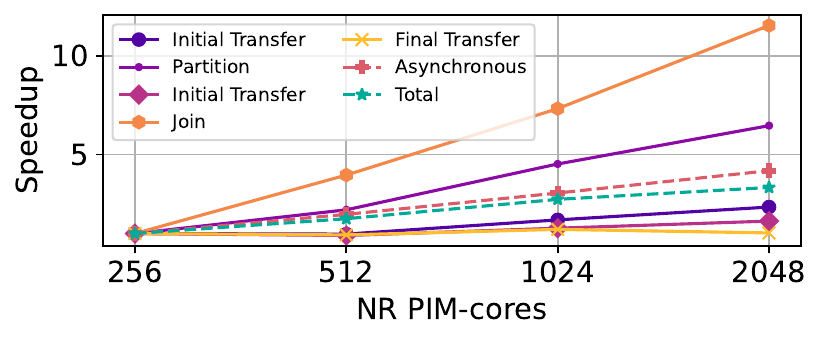}
\caption{Hash join operator}
\label{fig:hjoin_strong}
\end{subfigure}

\caption{Strong scaling comparison of operator implementations using execution time (bar plots) and speedup (line plots).}
\label{fig:strong_scaling}
\end{figure*}

Figure \ref{fig:strong_scaling} shows the strong scaling behavior (bar plots) and speedup (line plots), again broken down into the individual steps. As in the weak scaling analysis, we see the transfer times dominating.

For \texttt{selection} we see that the speedup of transfers is less than for the execution time (\ref{fig:sel_strong}). This is likely because the transfers are more bound by memory latency than bandwidth, since the size becomes relatively small.

The transfer times for sort (\ref{fig:saggr_strong}) and hash \texttt{aggregation} (\ref{fig:haggr_strong}) improve linearly with the number of PIM-cores. This is due to the host-PIM bandwidth increasing with the number of PIM ranks. The execution time scales faster than the transfer times for both operators, especially improving the performance of sort \texttt{aggregation}.

\recommendation{Using as many PIM-cores as possible not only improves kernel execution time, but also transfer times.}

\takeaway{Current PIM systems are still limited by the weaknesses of processor-centric architectures. This is because the processors have to be used for data transfers.}

There exist methods like broadcast joins \cite{BCJoin} to address the issue with communication in \texttt{join}. However, on PIM this only helps for very small table sizes (order of 10 MB), so we leave their evaluation to future work. A good solution to inter PIM-core communication will likely have to be implemented in hardware, as proposed in previous works \cite{DIMMLink23}.

\subsection{Operator comparison to CPU and GPU}

\begin{figure}[H]
    \centering
    \includegraphics[width=\linewidth]{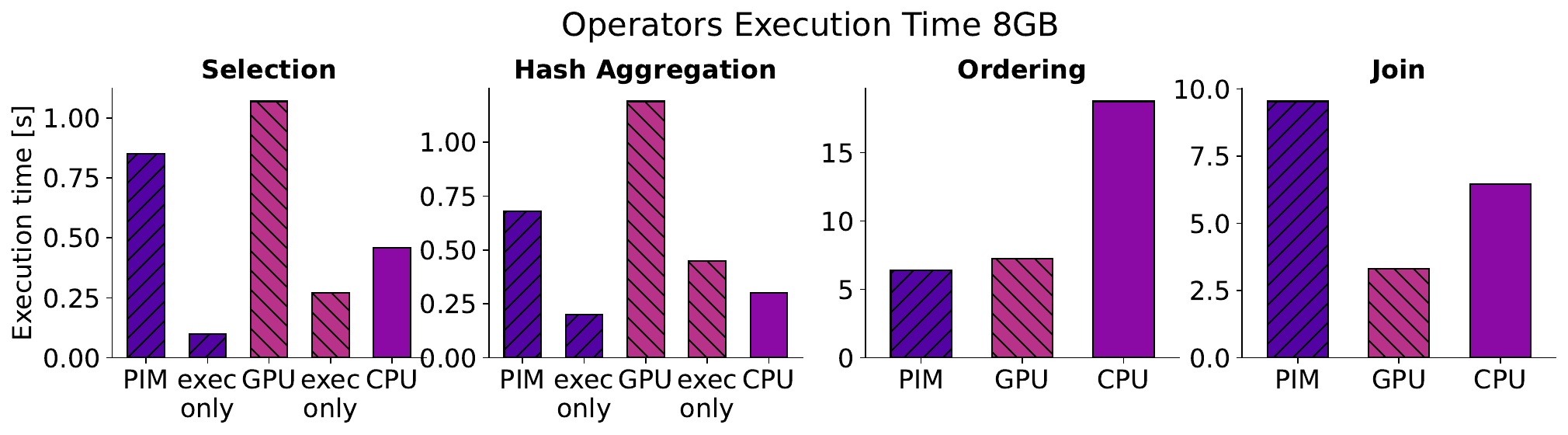}
    \caption{Operator execution time comparison to CPU and GPU for 8 GB data size.}
    \label{fig:comp_systems}
\end{figure}

Figure \ref{fig:comp_systems} compares the performance of the DB operators to the CPU and GPU. For all operators it shows the total execution time on all system including all transfers. For \texttt{selection} and \texttt{aggregation} it also shows the standalone execution time on PIM and GPU. We use 8 GB of data, since the GPU implementation cannot deal with bigger sizes.

Looking at the total time including transfers, we see that PIM and the GPU are slower than the CPU for \texttt{selection} and \texttt{aggregation}. This is because the complexity of these operators is comparatively small, similar to just transferring the data to the PIM system or GPU. In terms of just the execution time on the system, PIM is $4.5 \times$ faster than the CPU and $3.0 \times$ faster than the GPU for \texttt{selection}. For \texttt{aggregation} it is $1.5 \times$ and $2.2 \times$ faster respectively. This suggest that if we can avoid transfers, PIM performs considerably faster than the CPU.

\recommendation{Reusing data for multiple operators can improve the performance of DB operators on PIM by reducing transfers.}

For \texttt{ordering}, which has a higher complexity, PIM outperforms the CPU by $2.3 \times$ even when including all transfers. Compared to GPU it is slower with a speedup of $0.9 \times$, due to the data redistribution. For join it is slower than both CPU and GPU with a speedup of $0.6 \times$ and $0.3 \times$ respectively. With respect to the CPU this is caused by the small data size, as we show next.

We also compare the performance of the PIM system to the CPU using a bigger data size of 32GB. The overall performance including transfers, of \texttt{selection} and \texttt{aggregation}, worsens compared to CPU. The reason seems to be that the transfer times grow disproportionally. The speedup in standalone execution time remains the same for \texttt{selection}. Using the \textit{perf} \cite{perf} tool we measure an IPC of $1.2$ on the CPU. Relative to the maximum achievable, we estimate this to be about $40\%$, using the performance from the roofline model. We see that the PIM system is better utilized with $80\%$. The ratio of arithmetic performance to bandwidth in PIM systems better suits \texttt{selection} than the ratio in conventional CPUs. For \texttt{aggregation} the performance worsens on PIM with data size. However, the observed relative IPC is still higher on PIM than CPU with $90\%$ compared to $53\%$. The CPU profits from bigger cache sizes, while the PIM-cores content for the smaller SPM. Cache size is a key factor for the performance of \texttt{aggregation}.

For \texttt{ordering} and \texttt{join}, the execution time on the CPU increases significantly with data size, due to their complexity. PIM now outperforms it by $2.5 \times$ and $2.0 \times$ respectively. These operators move data around more often, which is where PIM can shine.

\begin{figure}[H]
    \centering
    \includegraphics[width=\linewidth]{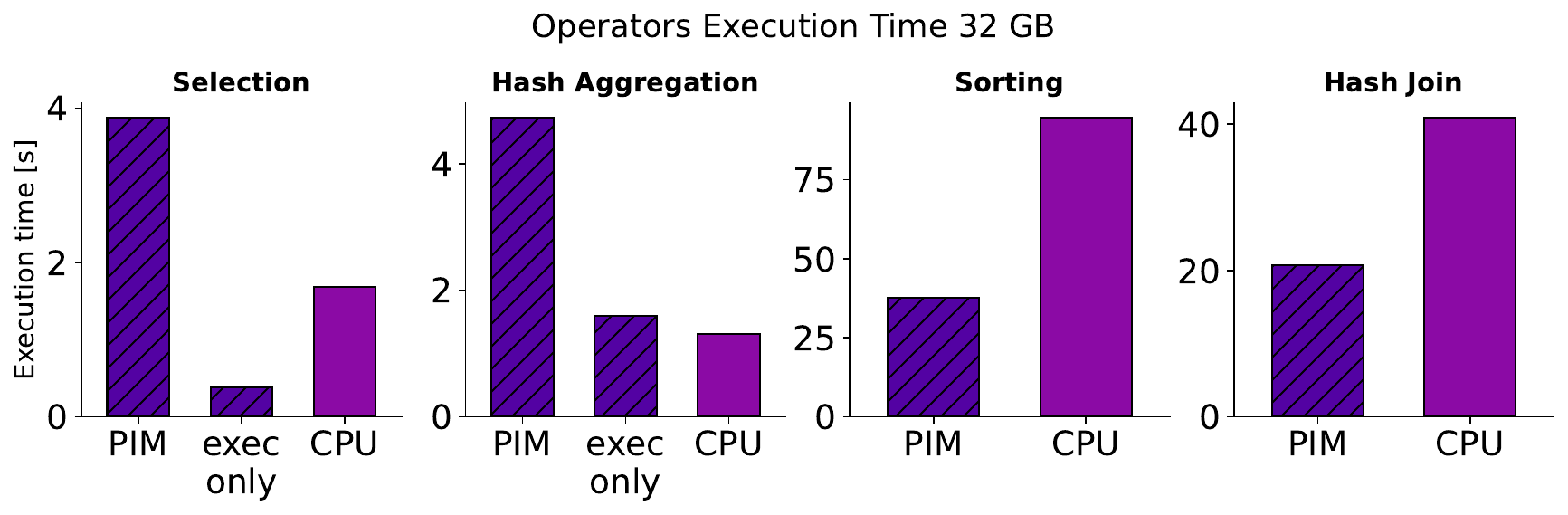}
    \caption{Operator execution time comparison to CPU for 32GB data size.}
    \label{fig:comp_cpu}
\end{figure}

\takeaway{Queries using \texttt{ordering} or \texttt{join} are better suited for acceleration using PIM systems compared to \texttt{selection} and \texttt{aggregation}.}

From our results for the individual operators we predict that PIM will perform well in queries with many operators and using \texttt{ordering} or \texttt{join}.

\subsection{TPC-H benchmarks}

\begin{figure}[H]
\centering
\includegraphics[width=1\linewidth]{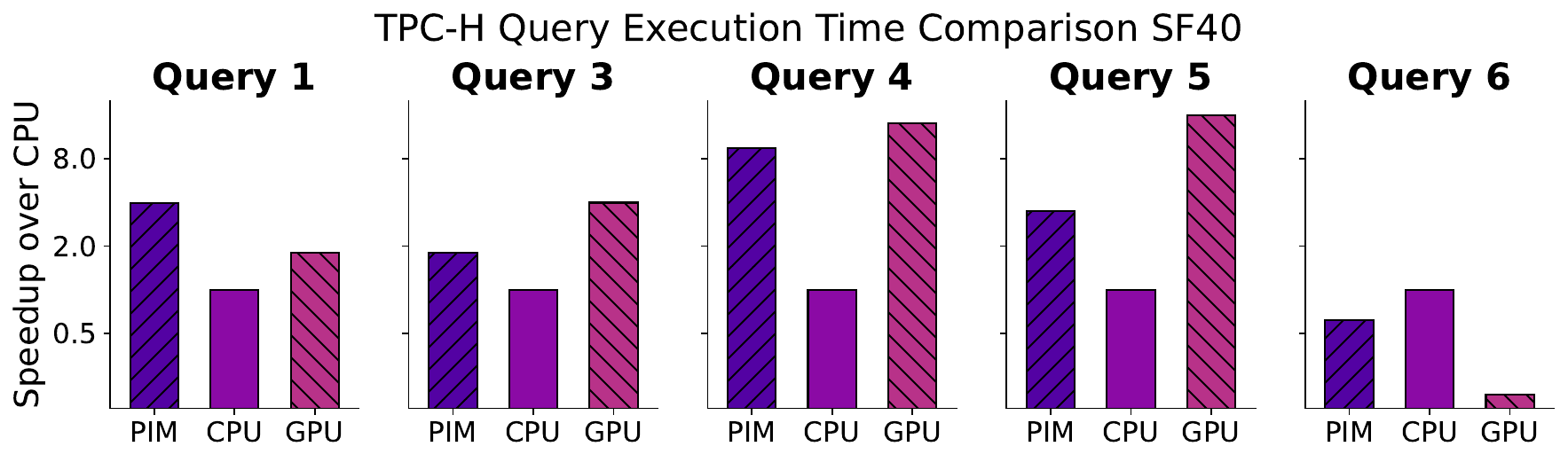}
\caption{Speedup of PIM and GPU compared to CPU in TPC-H queries.}
\label{fig:tpch_comp}
\end{figure}

Figure \ref{fig:tpch_comp} shows the speedup of PIM and GPU compared to CPU for five TPC-H queries. The PIM system outperforms the CPU in all queries expect for query 6. The average speedup compared to CPU is $3.9 \times$. PIM outperforms the GPU in queries 1 and 6.

Since queries 3, 4 and 5 depend on \texttt{join}, we can mainly explain the speedup over CPU with the speedup of this operation. Surprising is that the PIM system also outperforms the CPU in query 1, which only depends on \texttt{selection} and \texttt{aggregation}. The reason for this is that it has a very high selectivity and needs to calculate a lot of intermediate results. This incurs a high overhead on the CPU compared to the PIM system. Query 6 on the other hand, has a low selectivity and thus we observe a slowdown using PIM, like we do for the \texttt{selection} operator. The overall speedup is higher for the full queries, since the systems also have to move data that is not directly used for computation.

It is surprising that PIM can outperform the CPU even though the hardware support for arithmetic operations is limited. One reason is that most numbers in the TPC-H benchmark are fixed-precision decimals, which can be represented by 64 bit integers. The other reason is that the magnitude of the numbers are often small, which makes the software algorithms efficient. However, we also observe that the speedup of query 4 is the highest, which does not require multiplication. Even with the less efficient operations, the benefits of PIM outweigh them.

\observation{Analytical queries have properties which are suitable for PIM as our results of the TPC-H benchmark show.}

The GPU outperforms the PIM system in the queries that require \texttt{join}. This is expected, due to the data redistribution that needs to happen on the PIM system but not the GPU. The PIM system can outperform the GPU in the queries 1 and 6 that do not require joins, by $2.2 \times$ and $3.3 \times$ respectively.

Comparing both the GPU and PIM to the CPU performance, we can conclude that the limiting factor for the TPC-H queries is the main memory bandwidth. Compared to GPUs, the current UPMEM system cannot fully overcome this limitation, due to inter PIM-core communication. GPUs achieve high bandwidth while using unified memory. However, increasing this bandwidth is going to be more challenging than scaling up PIM. They have to increase the bus width, while in PIM it can increase with chip density.

\takeaway{Data analytics is well suited for more data-centric architectures as the results from PIM and also the GPU show.}

One limitation of PIMDAL is that it cannot deal with variable-length column fields. Unfortunately this is considerably more complex to implement compared to conventional systems, due to the memory management in the UPMEM SDK. However, this mainly relates to the ease of implementation, not the performance when using PIM. The main overhead would be in terms of memory transfers, which are comparatively cheap on PIM. Looking at the TPC-H queries with variable length data, the arithmetic intensity remains similar, suggesting our results will stay valid.

\section{Related Work}

\textbf{PIM hardware designs for database operators} create application specific hardware to accelerate them. The NON-VON architecture \cite{NonVon81} was an early attempt to create a system based on intelligent drives to accelerate database queries. Polynesia \cite{Polynesia21}, the mondrian data engine \cite{Mondrian17}, the Q100 \cite{Q10014}, JAFAR \cite{JAFAR15} co-design hardware and software for transactional and analytical query processing. These designs are all evaluated in simulation and are not commercially available, in contrast to the UPMEM system.

\textbf{Database operator designs on real-world PIM systems} are usually implemented on more general purpose architectures. AxDIMM \cite{AxDIMM22} runs the selection operator on a custom PIM architecture implemented on an FPGA that is not commercially available. PimDB \cite{PimDB23} implements selection and aggregation on the UPMEM system. However, it only evaluates them separately and does not implement joins. Another work \cite{Poseidon23} implements the graph database \textit{Poseidon} on the UPMEM system, to look at how multiple selection queries can be run concurrently and efficiently. Joins have also been implemented as a standalone operator on the UPMEM system in \textit{PID-Join} \cite{JoinDIMM23}. The work first implements the join operator and then proposes an improved method for data transfers. Our join implementation performs about $1.5\times$ faster than their unoptimized and about $2\times$ slower than their optimized join version. The speedup of \textit{PID-Join} relies on the improved transfer method, which is orthogonal to this work and could be used to further improve performance. \textit{SPID-Join}~\cite{SPIDJoin24} is an extension of \textit{PID-Join}, to deal with skews in the data distributions. None of these works implement all operators required to run full analytical queries. To our knowledge this is the first work that implements all operators and evaluates complete \textit{TPC-H} queries in order to fully demonstrate the capabilities of the UPMEM system for data analytics.

\textbf{In-memory database works} \cite{IMDBEval15, IMDBEval17, OracleIMDB15, MCIMDB13} look at the performance improvement by reducing the cost associated with accesses to persistent storage. The data is stored inside the main memory instead of loading it from disk. While this can improve the bandwidth over disk accesses, they are still bound by the main memory bottleneck. Here, we attempt to improve even on this aspect.

\section{Conclusion}

Using a data-centric architectures can significantly accelerate data analytics as we demonstrate with PIMDAL on the UPMEM PIM system. By measuring hardware metrics we can determine and evaluate the strengths and weaknesses of PIM for the implemented DB operators. Using five queries queries from the TPC-H benchmark, we show how PIM can achieve a speedup of $3.9\times$ compared to a CPU in analytical queries. However, we also find some key weaknesses in current PIM architectures, mainly the communication between PIM-cores. This will likely be the key point for future PIM systems to address. Overall we expect PIM to be able to accelerate complex analytical queries in the future, for example ones featuring ML as demonstrated in other PIM works~\cite{PIMML22}.

\section*{Acknowledgments}

We thank anonymous reviewers for feedback and the SAFARI group members for feedback and the stimulating intellectual environment they provide. We thank the UPMEM company for their technical support with this project, especially Sylvan Brocard, Julien Legriel and Denis Makoshenko. We acknowledge the generous gifts from our industrial partners, including Google, Huawei, Intel, and Microsoft. This work is supported in part by the ETH Future Computing Laboratory (EFCL), Semiconductor Research Corporation, AI Chip Center for Emerging Smart Systems (ACCESS), sponsored by InnoHK funding, Hong Kong SAR, and European Union’s Horizon programme for research and innovation [101047160 - BioPIM].

\bibliographystyle{unsrt}
\bibliography{refs}

\end{document}